\documentclass[manuscript,natbib]{aastex}

\usepackage{array}
\usepackage{multirow}
\newcolumntype{L}{>{\displaystyle}l}
\newcolumntype{C}{>{\displaystyle}c}
\newcolumntype{R}{>{\displaystyle}r}

\newcommand{\change}[1]{{\bf #1}}

\shorttitle{Detection of Lightning in PPD}
\shortauthors{Muranushi et al.}

\begin{document}

\title{Development of a Method for the Observation of Lightning in Protoplanetary Disks Using Ion Lines}

\author{Takayuki Muranushi}
\affil{RIKEN Advanced Institute for Computational Science,7-1-26, Minatojima-minami-machi, Chuo-ku, Kobe, Hyogo, 650-0047, Japan;takayuki.muranushi@riken.jp}

\author{Eiji Akiyama}
\affil{National Astronomical Observatory of Japan, 2-21-1 Osawa, Mitaka, Tokyo 181-8588, Japan; eiji.akiyama@nao.ac.jp}

\author{Shu-ichiro Inutsuka}
\affil{Nagoya University, Furo-cho,  Chikusa-ku, Nagoya, 464-8601, Japan; inutsuka@nagoya-u.ac.jp}

\author{Hideko Nomura} \affil{Department of Earth and Planetary Sciences, Tokyo Institute of Technology, nomura@geo.titech.ac.jp}

\and

\author{Satoshi Okuzumi} \affil{Department of Earth and Planetary Sciences, Tokyo Institute of Technology, okuzumi@geo.titech.ac.jp}

\begin{abstract}
{In this paper, we propose observational methods for detecting lightning in protoplanetary disks. We do so by calculating the critical electric field strength in the lightning matrix gas (LMG), the parts of the disk where the electric field is strong enough to cause lightning. That electric field accelerates multiple positive ion species to characteristic terminal velocities. In this paper, we present three distinct discharge models, with corresponding critical electric fields. We simulate the position-velocity diagrams and the integrated emission maps for the models. We calculate the measure of sensitivity values for detection of the models, and for distinguishing between the models. At the distance of TW-Hya (54pc), LMG that occupies $2\pi$ in azimuth and $25 \mathrm{au}<r<50 \mathrm{au}$ is  $1200\sigma$- to  $4000\sigma$-detectable. The \change{lower} limits of the radii of $5\sigma$-detectable LMG clumps are between 1.6 au and 5.3 au, depending on the models.}
\end{abstract}


\section{INTRODUCTION}
{Lightning in protoplanetary disks is one of the important topic in protoplanetary disk physics.
The existence of lightning is still an open question, and if it exists,
 it serves as one of the elementary electromagnetic processes, as one of the observational clue to measure the electromagnetic states of the disk, and as one of the candidate mechanism for chondrule heating.} {The observation data available today is huge, and open access to the observational results from the most advanced telescopes are available.} However, observation methods of the protoplanetary lightning using the advanced telescopes have not been seriously studied.

There has been a controversial debate on the existence and the mechanism of
protoplanetary disk lightning.
\citet{{doi:10.1006/icar.1997.5846}} argued that plasma conductivity is too large
for the lightning to take place.
\citet{{bibcode:1998A&A...331..121P}}
argued that unknown, efficient grain-grain charging process
is required to produce lightning.
Despite of these barriers, mechanisms that lead to lightning are proposed:
dust-dust collisional charging \citep{{bibcode:2000Icar..143...87D,bibcode:2010MNRAS.401.2641M}};
mutual positive feedback of thermal ionization and Joule heating \citep{{bibcode:2012ApJ...761...58H,bibcode:2013ApJ...767L...2M}}; electric field generated by magnetorotational instability (MRI)
 \citep{{doi:10.1086/432796,bibcode:2012ApJ...760...56M}}.
{Magnetized chondrules included in meteorites carry evidences of 500-1000G magnetic field during chondrule formation,
suggesting that they are struck by lightning
\citep{{bibcode:2000M&PS...35..537W}}.}

{Meanwhile, the understanding of the lightning ignition mechanism
have progressed in these twenty years.}
Attempts have been made to explain the mechanism that causes discharge at the point
well below the nominal dielectric strength of air
\citep{{doi:10.1029/JC076i024p05799,doi:10.1063/1.1656844}}. {As result,
new lightning models}
  such as Runaway breakdown
\citep{{doi:10.1016/0375-9601(92)90348-P,doi:10.1070/PU2001v044n11ABEH000939}}
have been proposed. {We adopt such progresses in understanding terrestrial lightning,}
  and propose a new observation method for
{the observational discrimination of the
protoplanetary disk lightning.
Observational studies searching for the disk lightning will contribute to the
understanding of the electromagnetic process in the protoplanetary disk,
and the source of chondrule heating.}

Lightning, or electrical breakdown is a result of large
electric field $E$. Electric field $E$ in protoplanetary disk can be generated by
magnetorotational instability (MRI) or by the collective motion of charged dust.
The breakdown model sets an upper limit $E \leq E_{\rm crit}$ to the electric field amplitude.
At the point $E = E_{\rm crit}$ electric discharge takes place, which increases the ionization degree of the medium
and prevents the further growth of the electric field amplitude.
Thus, electric field amplitude is kept under the upper limit ($E \leq E_{\rm crit}$) .

{
This electric field is a common feature of
the large volume of gas surrounding the lightning bolts.
In this study, we study
this electric-field feature as the possible predominant observational signals emitters.}

{Lightning bolt themselves are difficult to observe because}
 lightning is transient events, and typical radius of a lightning bolt is $5\times 10^3$ times
mean free path
\citep{{bibcode:1992ApJ...387..364P}}. This radius
is much smaller than the scale height of the disk. Hence
even if the critical condition is met, most of the time
most of the protoplanetary disk gas is in the region
outside the lightning bolts.
We call this lightning matrix gas (LMG).
Properties of LMG is no different from those of the disk gas without lightning,
but differ in one point that LMG is subject to critical
electric field $E \lesssim E_{\rm crit}$. In this paper, we explore the possible observational features of the LMG.

This paper is organized as follows. In section \ref{sec:Model}, we introduce the
discharge models, taking the Earth atposphere as an example (\S \ref{sec:DischargeModel}, \ref{sec:DischargeAir});
introduce the protoplanetary disk model (\S \ref{sec:DiskModel}); apply the discharge model to the disk gas
(\S \ref{sec:DiskDischargeModel}).
In section \ref{sec:Observation} we establish our observation model
by
{
\begin{enumerate}
\item calculating the terminal velocity of the ion molecules
\S \ref{sec:ObservationLines} ,
\item estimating the spectral irradiance
\S \ref{sec:ObservationEstimates} ,
\item constructing integral maps by radiative transfer simulations
\S \ref{sec:ObservationProfiles}.
\end{enumerate}
}

Given the simulated observational signals, we estimate the measure-of-sensitivity by
matched filtering (\S \ref{sec:MatchedFilter}).
Finally, in section \ref{sec:Observation}, we conclude and discuss the future directions of this research.
 \section{MODEL} \label{sec:Model} \subsection{Dielectric Strength of Air} \label{sec:DischargeModel} 
{We begin by} estimating the dielectric strength of the Earth atmosphere,
in order to introduce the reader to the discharge models we later apply to the
protoplanetary disk gas.

Dielectric strength of an insulating material is the maximum amplitude
of the electric field {that does not cause electric
breakdown in the material}. It is a physical property of central importance for the discharge
physics.
Lightning on Earth is {a} discharge phenomenon in the air. However, it has been a long standing
mystery that lightning takes place under electric field amplitude well below the
dielectric strength of the air {measured in the laboratory}.

The dielectric strength of {the} air at Normal Temperature and Pressure
(NTP; $20^{\circ}{\rm C}$ and 1atm)
are well established from laboratory experiments \citep{{isbn:9780028645865}}: \begin{eqnarray} E_0 = 30 {\rm kV/cm}. \label{NtpAirDielectricStrength}\end{eqnarray} The long-distance limit of Paschen's law states that the dielectric strength of gas
depends linearly on the gas number density \citep{{isbn:978-3-642-64760-4}}.
However, in the case of the Earth atmosphere the dependence of the dielectric strength on the number density is
known to be steeper than linear. This is explained by the effects of electron loss via three-body interactions
and also collisions to the water vapor molecules.
Empirical formulae are known
 \citep{{doi:10.1063/1.323084,isbn:9784130627184}}:
         \begin{eqnarray} E &=& E_0 \left( \frac{P}{P_0} \right) ^ {1.5 - 1.65} \label{TakahashiDischargeFormula} , \end{eqnarray} where $E_0$ and $P_0$ are the dielectric strength and the pressure of the air at ground level, respectively.
The formula predicts the dielectric strength of the air to be $17 {\rm kV/cm}$ and $10 {\rm kV/cm}$ at altitudes
3km and 6km, respectively.

On the other hand, intracloud lightning
is observed with electric field amplitude of
$140 {\rm V/cm}$ \citep{{doi:10.1029/96JD01625}} to
$150 {\rm V/cm}$ \citep{{doi:10.1029/JD091iD01p01231}}.
Cloud-to-ground lightning is observed with electric field amplitude of around
$1 {\rm kV/cm}$ \citep{{bibcode:1983JMSJ...61...656}} to
$2 {\rm kV/cm}$ \citep{{bibcode:1999JAtS...56.1561T}} .
  \subsection{Breakdown Models on Earth} \label{sec:DischargeAir} 
{In this section
   we introduce three breakdown models we are going to compare in this paper.
Note that in our formulation, the density functions of electrons and ions under
given electric field are well-understood (see e.g. \citet{{special:4874721036}}),
and thus we use the same density function for all the three breakdown models.
What we do not understand well is the dielectric strength ---
the amplitude of the electric field at which breakdown takes place.
The only difference among the three models is the assumed value of
the dielectric strength.
}
       \begin{itemize}\item[{\tt [T]} ]
In {\em Townsend breakdown model} the critical electric field
is such that an electron accelerated by the electric field
over its mean free path gains
kinetic energy large enough to ionize a neutral gas molecule.
It has widely been used in the meteorological context,
and also adopted into astrophysical context e.g. by
\citet{{doi:10.1006/icar.1999.6245,bibcode:2010MNRAS.401.2641M}} .
This model explains laboratory gas discharge experiments
(equation (\ref{NtpAirDielectricStrength})) well.

 \item[{\tt [DP]} ]
\citet{{doi:10.1103/RevModPhys.12.87}} has derived the formulae for equilibrium distribution
of electrons under constant electric field, neglecting the effects of inelastic
collision with atoms.
{When the electric field is weak, so that the work done by the electric field
       per mean free path $eEl_{\rm mfp}$ is much smaller than the electron kinetic energy,
the equilibrium distribution is nearly isotropic. The distribution is expressed as
the sum of the isotropic equilibrium and its first-order perturbation.
The average energy and the average velocity of the mean motion,
$\langle \epsilon \rangle$ and
$\langle v_z \rangle$, satisfy
$\langle \epsilon \rangle =0.43 eEl_{\rm mfp} \sqrt{M/m_e}$ and
$\langle v_z \rangle = 0.9 \sqrt{eEl_{\rm mfp}} (m_e M)^{ -1/4}$, respectively.
Here, $m_e$ and $M$ are the masses of the electron and the collision partner, respectively.
}

{\em Druyversteyn-Penning (DP) breakdown model} assumes that the breakdown
takes place when $\langle \epsilon \rangle$ exceeds the ionization energy.
{ Since
 the factor $\sqrt{M/m_e}$ makes
         the average energy $\langle \epsilon \rangle$ in DP {breakdown} model nearly 100 times larger than that
 in Townsend breakdown model,
DP {breakdown} model allows for breakdown under electric field amplitude at about $10^{ -2}$ times that
of the Townsend {breakdown} model.}
The model is introduced as a protoplanetary disk lightning model by
\citet{{doi:10.1086/432796}}.

  \item[{\tt [R]} ]
\citet{{doi:10.1016/0375-9601(92)90348-P}}
have proposed {the} {\em runaway breakdown model}
and
\citet{{doi:10.1070/PU2001v044n11ABEH000939}}
provided a detailed review of the model.
{ In this model, the equilibrium of the electrons with
 relativistic ($\sim 1$MeV) kinetic energy, much larger than the average of  Maxwellian energy distribution, plays an important role.}
{Because the ionization losses for electrons is inversely proportional to the kinetic energy in the non-relativistic regime, }
the mean free path for such fast electrons is much longer than that for thermal
electrons.
{In the runaway breakdown model, the exponential growth of the number of relativistic electrons takes place, once
the electric field is large enough to balance the ionization losses for certain energy range of the relativistic electrons.
(We define that the acceleration criteria is met for those electrons.)
The ionization processes generate
spectrum of fast and slow electrons. Fast electrons that meets the acceleration criteria contribute to the exponential growth, while the slow electrons are large in number, and increase the ionization degree of the matter and ultimately lead to the electric breakdown of the matter.
}
{Thus, }
  runaway breakdown can take place at an electric field value much weaker than that of a Townsend breakdown.
{The} runaway breakdown model better explains the lightning observations in the Earth{'s} atmosphere and is used as the discharge
model in thunderstorm simulations studies,
e.g. by \citet{{bibcode:2002JGRD..107.4075M}}. \end{itemize}
In order to estimate the dielectric strength of gas,
we need to compute {the} energy distribution of electrons.
Since the interactions of electrons with even the simplest atoms and molecules
have profound details \citep{{isbn:3-540-64296-X,isbn:3540653473,isbn:354044338X}},
this requires difficult numerical computations \citep{{doi:10.1063/1.329081}}.
In this paper, we will instead resort to a simple calculation that reproduces
the observed values from the discharge models.

First, we derive the dielectric strength of air at ground level from {the} Townsend model.
Air consists of
78\% $\rm N_2$,  21\% $\rm O_2$, and 1\% $\rm Ar$ (volume fractions).
Air number density at NTP is $2.504 \times 10^{19} {\rm cm^{ -3}}$ .
The ionization energy of these chemical species are
$\Delta W_{\rm N_2} = 15.6~{\rm eV} $,
$\Delta W_{\rm O_2} = 12.1~{\rm eV}$, and
$\Delta W_{\rm Ar}  = 15.8~{\rm eV}$, respectively.
Of these $\Delta W_{\rm O_2} \sim 12 {\rm eV}$ is the smallest, so we estimate
the electric field amplitude $E_{\rm crit}$ required to accelerate the electron
{up to} 12eV; i.e. we solve $12 {\rm eV} = e E_{\rm crit} l_{\rm mfp}$.
The inelastic collisional cross sections ($\sigma_{\rm inel}$) of  $\rm N_2, O_2, Ar$ for 12 eV electrons are
$0.8, ~1.8, ~0.0 \times 10^{ -16} {\rm cm^{ -2}}$
\citep{{isbn:3-540-64296-X,isbn:354044338X}}.
Therefore, the mean inelastic cross section of air for 12eV electrons is
$1.0 \times 10^{-16} {\rm cm^{ -2}}$.
Therefore, $l_{\rm mfp} = (n_n \sigma_{inel})^{ -1 } = 4.0 \times 10^{-4} {\rm cm}$.
This gives
\begin{eqnarray}
E_{\rm crit} &=& 30 {\rm kV/cm},
\end{eqnarray}
 which is in agreement with the dielectric strength of air at ground level
(equation (\ref{NtpAirDielectricStrength})).

On the other hand, according to Druyversteyn-Penning model
, average kinetic energy of electron
under the electric field $E$ is \citep{{doi:10.1086/432796}}
\begin{eqnarray}
  \langle \epsilon \rangle = 0.43 e E l_{\rm mfp} \sqrt{\frac{M}{m_e}},
\end{eqnarray}
where $M$ is the mass of the collision partner,
and the dielectric strength $E_{\rm crit}$ is the solution of $\langle \epsilon \rangle = \Delta W$.
In the case of the air at NTP,
since mean molecular weight of air is 28.96~{\rm g/mol},
 $M = 4.81 \times 10^{-23}~{\rm g}$. 
Note that $l_{\rm mfp}$ in Druyversteyn-Penning model means elastic
mean free path
$l_{\rm mfp} = (n_n \sigma_{el})^{ -1} = 3.59 \times 10^{-5} {\rm cm}$.
$l_{\rm mfp}$ is calculated from elastic cross sections of the elemental molecules at 12eV
($\sigma_{el} =
 1.16 \times 10^{-15}~{\rm cm^2},
 9.00 \times 10^{-16}~{\rm cm^2},
 1.74 \times 10^{-15}~{\rm cm^2}$,
respectively, for $\rm N_2, O_2, Ar$, see \citet{{isbn:3-540-64296-X,isbn:354044338X}}.)
Therefore,
$E_{\rm crit} = 3.38~{\rm kV/cm} $.

Finally, according to the runaway breakdown model the dielectric strength
$E_{\rm crit}$ is the electric field amplitude where
the acceleration by the electric field balances
the ionization loss for minimum ionizing electrons.
Minimum ionizing electrons are electrons with such kinetic energy $\varepsilon$
that  for them the ionization loss is the smallest.
{The kinetic energy of the minimum ionizing electrons is about $1$MeV, where ionization loss
is the dominant energy sink for the electrons
\citep{{doi:10.1070/PU2001v044n11ABEH000939}}
.}
The ionization loss of an electron {per unit time} as a function of $\varepsilon$
is formalized by
\citet{{doi:10.1002/andp.19303970303,doi:10.1007/BF01342532,doi:10.1007/BF01344553}}.
We use the following form of Bethe formula from
\citet[chap 5.5]{{isbn:978-0-521-75618-1}}:
   \begin{eqnarray} -\frac{d\varepsilon}{dx} &=&
\frac{e^4 {\bar Z} n_n}{8 \pi {\epsilon_0}^2 m_e c^2}a(\gamma), \\
\mathrm{where}~~~a(\gamma) &=& \left(\frac{c}{v}\right)^2
   \left[ \ln\frac{\gamma^3 {m_e}^2 v^4 }{2 (1+\gamma){\bar I}^2}
 - \left(\frac{2}{\gamma} - \frac{1}{\gamma^2}\right)\ln 2
 + \frac{1}{\gamma^2}
 + \frac{1}{8}\left(1 - \frac{1}{\gamma}\right)^2\right].
    \end{eqnarray}
Here,
$ \gamma = (1 - v^2/c^2)^{ -1/2} $ is the Lorentz factor of the electron,
$\varepsilon = (\gamma - 1) m_e c^2$ is the electron kinetic energy,
${\bar Z} n_n$ is the number density of ambient electrons of the matter.
$\bar I$ is the mean excitation energy, a parameter to be fitted to
laboratory experimental data. We use the value of
$\bar I_{\rm air} = 86.3 {\rm eV}$ from ESTAR database \citep{{special:nist-estar}}.

For the case of the air $a(\gamma)$ takes its minimum
$a_{\rm min} = 20.2$ at $\gamma = 3.89$
or $\varepsilon = 1.48 {\rm MeV}$. The dielectric strength $E_{\rm crit}$
is the solution of the following work-balance equation
\begin{eqnarray}
  eE-\frac{d\varepsilon}{dx} &=& 0,
\end{eqnarray}
which is
\begin{eqnarray}
E_{\rm crit}
&=& \frac{e^3 {\bar Z} n_n}{8 \pi {\epsilon_0}^2 m_e c^2}a_{\rm min}, \nonumber \\
&=&  1.9~{\rm kV/cm} .
\end{eqnarray}
   
Here we summarize the three models. The dielectric strength of the gas is proportional to
the number density of the gas. It is this proportional relation that leads to {the} constant ion velocity
we present in this paper.

\begin{eqnarray}
\begin{array}{CCCCC}
E_{\rm c, T} &=& \frac{\Delta W}{e} \sigma_{\mathrm{ inel}}  n_n&=&
 30.1~{\rm kV/cm} \cdot
 \left( \frac{n_n}{n_{0,\mathrm{air}}}  \right)^{1}   , \\
E_{\rm c,DP} &=& \frac{\Delta W}{0.43} \sqrt{\frac{m_e}{M}} \sigma_{\mathrm{el}} n_n  &=&
 3.4~{\rm kV/cm} \cdot
 \left( \frac{n_n}{n_{0,\mathrm{air}}}  \right)^{1}  , \\
E_{\rm c, R} &=& \frac{e^3 a_{\rm min} {\bar Z} }{8 \pi \epsilon_0 m c^2} n_n&=&
 1.9~{\rm kV/cm} \cdot
 \left( \frac{n_n}{n_{0,\mathrm{air}}}  \right)^{1}   .
\end{array} \label{eq:DischargeAir}
\end{eqnarray}
  \subsection{The Disk Model} \label{sec:DiskModel} The minimum-mass solar nebula (MMSN) model \citep{{bibcode:1981PThPS..70...35H}} has been widely used in studies of the protoplanetary disk, with fruitful results.
    Recent observations have contributed to sophistication of the disk models, and {have also}  reported qualitative values
    for the inner and outer edge radius of the disks
    \citep{{bibcode:2002ApJ...581..357K,bibcode:2009ApJ...700.1502A,bibcode:2010ApJ...723.1241A,doi:10.1146/annurev-astro-081710-102548}}.
    However, such observational values  for
    geometry and mass of
    specific objects still contain uncertainty factors of 2-3, and are subjects of debate.
    See e.g.
    \citet{{arxiv:1402.6597}}.

    Some of the features common to the recent models are that the power-law indices
    of the surface density distribution is
    close to 1 rather than 1.5, and that there are exponential cut-off at the outer
    edge of the disk
    Therefore, we use simple model proposed by
    \citet{{doi:10.1093/pasj/65.6.123}} that captures these common features,
    and adopt the values of TW Hya reported by
    \citet{{bibcode:2002ApJ...568.1008C}}. Our disk model is as follows:

\begin{eqnarray}
\Sigma{}\left(r\right)&=& 6.4 \times 10^{2}
\left(\frac{r}{1\mathrm{au}}\right)^{ -1}
\exp\left(-\frac{3r}{r_{\rm out}}\right)
\mathrm{g/cm^{2}}
\hspace{1cm} {\rm for}~ ~  r>r_{\rm in}, \nonumber
\\
\Sigma{}\left(r\right)&=&0\hspace{1cm} {\rm otherwise} ,
\\
T\left(r\right)&=& 273
\left(\frac{r}{1\mathrm{au}}\right)^{ -\frac{1}{2}}\mathrm{K} .
\end{eqnarray}

Here, $r_{\rm out} = 150 {\rm au}$ is the outer radius of our model disk.
We also introduce an inner cutoff at
 $r_{\rm in} = 3.5 {\rm au}$.
           The assumption of the hydrostatic equilibrium leads to the vertical distribution of the gas
\begin{eqnarray}
\rho(r,z)&=&\rho_0(r) \exp \left(-\frac{z^2}{2h^2}\right) \nonumber \\
&=&  5.08 \times 10^{-10}
\left(\frac{r}{1\mathrm{au}}\right)^{ -\frac{3}{2}}
\exp \left(-\frac{z^2}{2h^2}\right) {\mathrm{g ~ cm^{ -3 }}}  ,\\
\mathrm{where} ~~~
h(r) &=& \frac{c_s}{\Omega} \nonumber \\
&=&   3.29 \times 10^{-2}
\left(\frac{r}{1\mathrm{au}}\right)^{ \frac{5}{4}}  {\mathrm{au}} ,
 \\
c_s(r) &=& \sqrt{\frac{k_B T(r)}{\mu m_p}},\\
\Omega_K(r) &=& \sqrt{\frac{G M_{\odot}}{r^3}},\\
v_K(r)     &=& \sqrt{\frac{G M_{\odot}}{r}}.
\end{eqnarray}
{Here $\mu m_p$ is the mean molecular mass of the gas.}
    \change{Therefore the number density of $\mathrm{H_2}$ is
\begin{eqnarray}
n_{\mathrm{H_2}}(r,z)
&=&  1.52 \times 10^{14}
\exp \left(-\frac{z^2}{2h^2}\right)
\frac{\Sigma(r)}{\Sigma(1 \mathrm{au})} {\mathrm{cm^{ -3 }}} .
\end{eqnarray}}
   \subsection{Breakdown Models on Protoplanetary Disks} \label{sec:DiskDischargeModel} 
Here we estimate the dielectric strength of the protoplanetary disk gas.
\change{In our disk model the gas density at the equatorial plane, $r = 1{\rm au}$ is
$n_{0,\mathrm{ppd}} = 1.52 \times 10^{14}
{\rm cm^{ -3}}$.}
We assume that protoplanetary disk gas consists of
$\rm H_2,He,CO,O_2$ and their volume fractions are
$0.92, 7.8\times10^{ -2},2.3\times10^{ -4},1.3\times10^{ -4}$, respectively
\citep[chap. 3.4.6.]{{bibcode:2009LanB...4B...44L}} .
We used cross sections data for 15eV electrons
tabulated in   \citep{{isbn:3-540-64296-X,isbn:354044338X}} (c.f. Table \ref{tbl:CrossSection}),
{since}
 $\Delta W_{\rm H_2} = 15.43{\rm eV}$ and 15eV is the closest table index that is found in the database.

\begin{table}[t]
\begin{center}
\bf
  \begin{tabular} {|c|c|c|}
\hline
species & $\sigma_{\rm inel}$ & $\sigma_{\rm el}$ \\
\hline
$\rm H_2$  &  $1.6 \times 10^{-16} {\rm cm}^{ -2} $ &  $6.6 \times 10^{-16} {\rm cm}^{ -2} $\\
$\rm He$   &  $0.0 {\rm cm}^{ -2} $ &  $3.6 \times 10^{-16} {\rm cm}^{ -2} $\\
$\rm CO$   &  $5.1 \times 10^{-18} {\rm cm}^{ -2} $ &  $1.1 \times 10^{-15} {\rm cm}^{ -2} $\\
$\rm O_2$  &  $1.8 \times 10^{-16} {\rm cm}^{ -2} $ &  $8.9 \times 10^{-16} {\rm cm}^{ -2} $\\
\hline
\end{tabular}
\end{center}
\caption{Collisional cross sections of the molecules for 15eV electrons.}
\label{tbl:CrossSection}
\end{table}

 Calculations similar to those in the previous section {lead} to the following values:

\begin{eqnarray}
\begin{array}{CCCCC}
E_{\rm c, T} &=& \frac{\Delta W}{e} (\sigma_{\mathrm inel})  n_n&=&
 3.0 \times 10^{-1} \left( \frac{n_n}{n_{0,\mathrm{ppd}}}  \right) \mathrm{V/cm}  , \\
E_{\rm c,DP} &=& \frac{\Delta W}{0.43} \sqrt{\frac{m_e}{M}} \sigma_{\mathrm el} n_n  &=&
 5.0 \times 10^{-2} \left( \frac{n_n}{n_{0,\mathrm{ppd}}}  \right) \mathrm{V/cm}  , \\
E_{\rm c, R} &=& \frac{e^3 a_{\rm min} {\bar Z} }{8 \pi \epsilon_0 m c^2} n_n&=&
 1.4 \times 10^{-3} \left( \frac{n_n}{n_{0,\mathrm{ppd}}}  \right) \mathrm{V/cm}  .
\end{array} \label{eq:DischargeDisk}
\end{eqnarray}

\section{OBSERVATION} \label{sec:Observation} \subsection{Calculation of the Terminal Velocity of the Ions} \label{sec:ObservationLines} 
  The goal of this section is to calculate the Doppler broadening of the molecular ion lines in the disk,
  which reflects the electric field strength in the protoplanetary disk.
  In order to establish the observation procedure, we calculate the collisional cross sections and the terminal
  velocities of the molecules. Then, we can estimate the optical depths and the spectral irradiances of the specific
  lines. We simulate the observational images using the calculated spectral irradiances. Finally, we establish a model discrimination procedure based on matched-filtering.

  We choose three ion species: $\mathrm{HCO}^{+}$,  $\mathrm{DCO}^{+}$ and  $\mathrm{N_2H}^{+}$  lines, whose observations have been performed
  \citep{{bibcode:2011ApJ...734...98O,bibcode:2010ApJ...720..480O}} .
Such charged chemical species are accelerated upto their respective terminal velocity by
the electric field of the LMG.
Let  $\varepsilon_I$ be the kinetic energy of a particle of such an ion species $I$.
{We can calculate the value of $\varepsilon_I$ at the equilibrium by solving
\begin{eqnarray}
\kappa_{I,n} \varepsilon_I = {e E_{\rm crit} l_{{\rm mfp},I}},
\end{eqnarray}

Here,
$\kappa_{I,n} = \frac{2m_I m_n}{(m_I+m_n)^2}$
is the fraction of ion energy loss per collision, and
$m_I$ and $m_n$ are the masses of the ion and the neutral molecules, respectively
\citep{{special:4874721036}}.
}

As shown in equations (\ref{eq:DischargeAir}) and (\ref{eq:DischargeDisk}),
the dielectric strength $E_{\rm crit}$ is proportional to the gas number density $n_n$.
Let $A$ be the proportionality factor and $E_{\rm crit} = A n_n$ .
Now, the mean free path $l_{{\rm mfp},I} = 1/\sigma_I(\varepsilon_I) {n_n}$ is inversely proportional to the gas number density $n_n$.
This means that the obtained kinetic energy $\varepsilon_I$ is independent of the gas number density.

\begin{eqnarray*}
 \varepsilon_I &=& e E_{\rm crit} l_{{\rm mfp},I} (\kappa_{I,n})^{ -1} \\
 &=& \frac {e A (m_I+m_n)^2 } { 2 \sigma_I(\varepsilon_I) m_I m_n }
\end{eqnarray*}

{ The value of $A$ only depends} on the lightning model,
so {it} is universally the same  in a protoplanetary disk.
This \change{feature is what we propose as a new signal} of the lightning models in the disk.
\change{ Recall that the proportionality factor $A$ for the three lightning models are given as $E_{\rm crit} = A n_n$
in Equations (\ref{eq:DischargeDisk}).
 }
The predicted
$\varepsilon_I$
and the velocities of the ion species {are shown in}
Table \ref{tbl:terminalVelocity}.
In the Appendix we describe
the detail
of the cross section model we have used in order to estimate the above cross sections.

\begin{table}[t]

 {

 \begin{center}
  Cross Section Model XL

  \begin{tabular} {|c|ccc|}
\hline
 &  $\mathrm{HCO}^{+}$ & $\mathrm{DCO}^{+}$ & $\mathrm{N_2H}^{+}$ \\
\hline
T
  & $8.6 \times 10^{4}$ cm/s
  & $8.6 \times 10^{4}$ cm/s
  & $8.6 \times 10^{4}$ cm/s
  \\
DP
  & $2.5 \times 10^{4}$ cm/s
  & $2.5 \times 10^{4}$ cm/s
  & $2.5 \times 10^{4}$ cm/s
  \\
R
  & $2.2 \times 10^{3}$ cm/s
  & $2.2 \times 10^{3}$ cm/s
  & $2.2 \times 10^{3}$ cm/s
  \\
\hline
\end{tabular}
\end{center}
}

\begin{center}
  Cross Section Model XM

  \begin{tabular} {|c|ccc|}
\hline
 &  $\mathrm{HCO}^{+}$ & $\mathrm{DCO}^{+}$ & $\mathrm{N_2H}^{+}$ \\
\hline
T
  & $4.2 \times 10^{5}$ { cm/s}
  & $4.2 \times 10^{5}$ { cm/s}
  & $4.2 \times 10^{5}$ { cm/s}
  \\
DP
  & $1.2 \times 10^{5}$ { cm/s}
  & $1.2 \times 10^{5}$ { cm/s}
  & $1.2 \times 10^{5}$ { cm/s}
  \\
R
  & $1.1 \times 10^{4}$ { cm/s}
  & $1.1 \times 10^{4}$ { cm/s}
  & $1.1 \times 10^{4}$ { cm/s}
  \\
\hline
\end{tabular}
\end{center}

{

\begin{center}
  Cross Section Model XS

  \begin{tabular} {|c|ccc|}
\hline
 &  $\mathrm{HCO}^{+}$ & $\mathrm{DCO}^{+}$ & $\mathrm{N_2H}^{+}$ \\
\hline
T
  & $2.0 \times 10^{6}$ cm/s
  & $2.0 \times 10^{6}$ cm/s
  & $2.0 \times 10^{6}$ cm/s
  \\
DP
  & $5.9 \times 10^{5}$ cm/s
  & $5.9 \times 10^{5}$ cm/s
  & $5.9 \times 10^{5}$ cm/s
  \\
R
  & $5.2 \times 10^{4}$ cm/s
  & $5.2 \times 10^{4}$ cm/s
  & $5.2 \times 10^{4}$ cm/s
  \\
\hline
\end{tabular}
\end{center}
}

 \caption{
{
The terminal velocities of the molecular ions, for cross section models
XL, XM and XS. The terminal velocities are the characteristic features we use
for observational measurement of the dielectric strength.}
}\label{tbl:terminalVelocity}
\end{table}

   \subsection{Estimation of the Observational Signals}
   \label{sec:ObservationEstimates}

  The column density corresponding to optical depth $\tau_{\nu 0}=1$ is estimated as follows
  \citep[see the appendix of][]{{bibcode:1986ApJ...303..416S}}:

  \begin{eqnarray}
N_{\tau_{\nu 0}=1} =
\frac{3 \epsilon_0 k_B T_{\rm ex} \Delta v_{\rm gas}} {2 \pi^2 B \mu^2 \cos \theta}
\frac{1}{(J+1)}
\exp
\left(
  \frac{hBJ(J+1)}{k_B T_{\rm ex}}
\right)
\left(
  1 - \exp
  \left(
    - \frac{h \nu_0}{k_B T_{\rm ex}}
  \right)
\right),
        \end{eqnarray}

  where $T_\mathrm{ex}$ is the excitation temperature,
  $\Delta v_\mathrm{gas}$ is the Doppler broadening of the target molecule,
  $B$ is the rotational constant of the molecule, $\mu$ is its electric dipole matrix element,
  $\theta$ is the angle between the disk axis and the line of sight,
  $h \nu_0$ is the energy difference between the two levels
  and $J$ is the rotational quantum number of the lower state.

  The optical depth of the disk with column density $N$ is

\begin{eqnarray}
\tau_{\nu_0}(N) = N / N_{\tau_{\nu 0}=1} ,
\end{eqnarray}

  so that the intensity is

\begin{eqnarray}
I(\nu_0) = B(\nu_0,T) \left(1-\exp (-\tau_{\nu_0}(N))\right)   .
\end{eqnarray}

  Here $ B(\nu_0,T) $ is the vacuum brightness of a black body at frequency $\nu_0$
  (see \citet[chap. 2.7]{{isbn:3-540-29692-1}}).

The spectral irradiance of the disk $E(\nu)$
as a function of $\nu$ is

\begin{eqnarray}
E(\nu) &=& \frac{1}{D^2}
\int \int I(\nu_0,r) \exp
\left(
- \frac{m c^2 d(\nu; \nu_0,r)^2}{2 k_B T {\nu_0}^2}
\right)
r \mathit{dr} \mathit{d\varphi} \cos \theta , \label{eq:SpectralIrradiance} \\
\mathrm {where} ~~~
d(\nu; \nu_0,r) &=& \nu - \nu_0 - \frac{v_K(r)}{c} \cos \varphi \sin \theta \nonumber .
\end{eqnarray}

The integral is done in cylindrical coordinates $(r,\phi)$, and $\theta$ is the inclination angle of the disk.

   We consider $\rm HCO^{+}~3-2$, $\rm DCO^{+}~3-2$ and  $\rm N_2H^{+}~3-2$ lines.
   Their frequencies are
   $267.56$GHz,
   $216.12$GHz and
   $279.52$GHz, respectively.
   At 100au of the model disk $T =  27 {\rm K}$.
   
   For simplicity we assume that fractional abundances of the ion species are uniform
   within the disk. The population of the ionized species can be drastically increased
   or decreased as a result of lightning, but to estimate such population is beyond the
   scope of this paper.
   We adopt the XR+UV-new chemical process model of
   \citet{{doi:10.1088/0004-637X/747/2/114}},
   and
   {use the abundance values at  $r=$100au, $z=3h$, since the electric field is most
   likely to reach the critical amplitude at higher altitudes of the disk
    \citep{{bibcode:2012ApJ...760...56M}}.
    Thus, we assume that the fractional abundances (relative to $\rm H_2$) of $\rm N_2H^{+}$ is
    $5.3 \times 10^{-10} $.     We assume the fractional abundance of $\rm HCO^{+}$ and $\rm DCO^{+}$ to be
    $9.0 \times 10^{-9} $ and
   $3.0 \times 10^{-9}$, respectively, based on  paper by
   \citet{{bibcode:2013A&A...557A.132M}} that reports observation of  enhancement in DCO abundance.}
   Therefore, the column densities of $\rm HCO^{+}$, $\rm DCO^{+}$ and $\rm N_2H^{+}$
   are $1.8 \times 10^{15} {\rm cm^{ -2}}$,
   $6.0 \times 10^{14} {\rm cm^{ -2}}$,
   and $1.1 \times 10^{14} {\rm cm^{ -2}}$,
   respectively.

   Assuming that there is no lightning and that the molecules are in their thermal velocities,
   $N_{\tau_{\nu 0}=1}$ for the three lines are
   $7.78 \times 10^{9} ~{\rm cm^{ -2}}$,
   $6.74 \times 10^{9} ~{\rm cm^{ -2}}$ and
   $1.39 \times 10^{10} ~{\rm cm^{ -2}}$, respectively.
   On the other hand, $N_{\tau_{\nu 0}=1}$ for the three lines are
   $6.15 \times 10^{11} ~{\rm cm^{ -2}}$,
   $5.42 \times 10^{11} ~{\rm cm^{ -2}}$ and
   $1.10 \times 10^{12} ~{\rm cm^{ -2}}$, respectively,
   if the molecules are accelerated by the lightning electric field.

   We can see that the disk is optically thick for all of the lines at 100au.
   However,
   all the three lines {become} two degrees of magnitude more transparent under the effect of the
   critical electric field. This is a result of the molecular speed becoming faster.
   Consequently, observed line profiles are broadened. This Doppler broadening of the lines of the charged molecular
   species are the key observational features to observe the characteristic speed of the molecules, and therefore
   the electric field strength in the protoplanetary disk.

   \subsection{Calculations of the Line Profiles and Integral Maps by Radiative Transfer}
   \label{sec:ObservationProfiles}

   \begin{table}\begin{center}
   \begin{tabular}{|c|c|c|}
     \hline
     disk model name & discharge & LMG region \\
     \hline
     N   & no discharge  & \\
     T25 & Townsend discharge & $25{\mathrm{au}} < r < 50{\mathrm{au}}$\\
     T50 & Townsend discharge& $50{\mathrm{au}} < r < 100{\mathrm{au}}$\\
     DP25 & Druyversteyn-Penning discharge & $25{\mathrm{au}} < r < 50{\mathrm{au}}$\\
     DP50& Druyversteyn-Penning discharge & $50{\mathrm{au}} < r < 100{\mathrm{au}}$\\
     R25 & runaway dischage & $25{\mathrm{au}} < r < 50{\mathrm{au}}$\\
     R50& runaway dischage & $50{\mathrm{au}} < r < 100{\mathrm{au}}$\\
     \hline
   \end{tabular}
   \end{center}
  \caption{Our seven disk models, their respective discharge models and LMG distribution models.}\label{tbl:DiskModels}
  \end{table}

   \begin{figure}
   \includegraphics[angle=270,width=8cm]{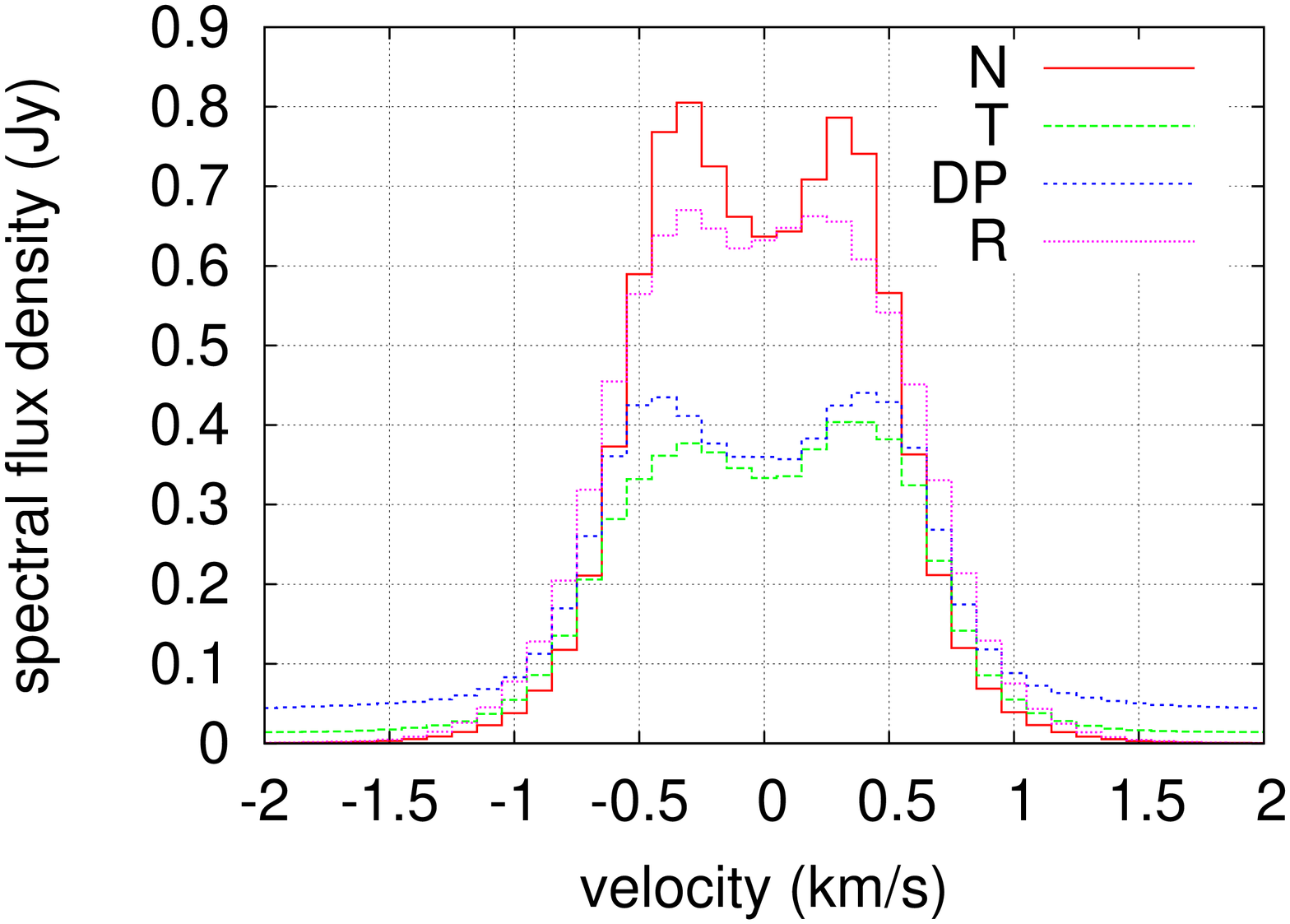}
   \includegraphics[angle=270,width=8cm]{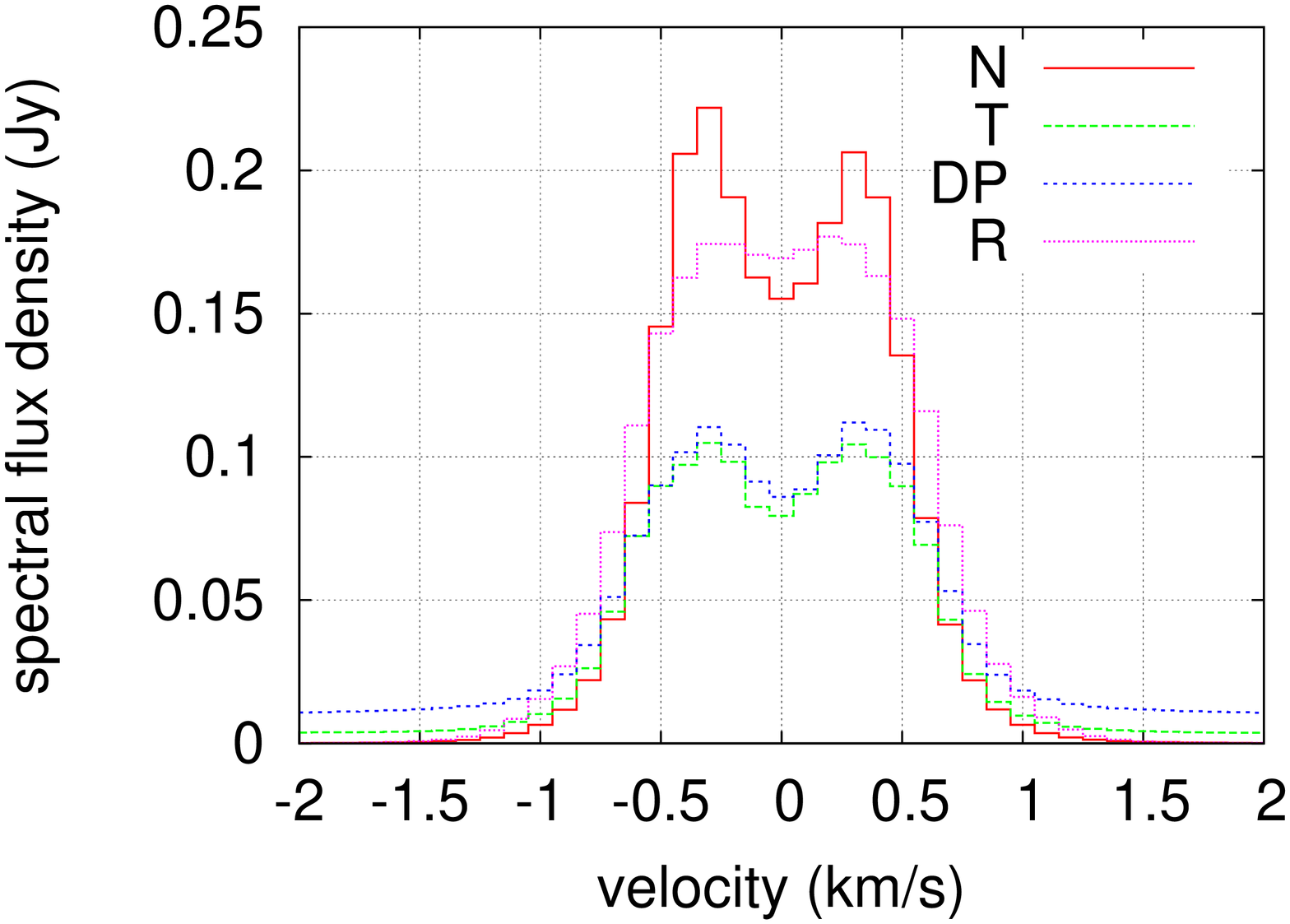}
   \includegraphics[angle=270,width=8cm]{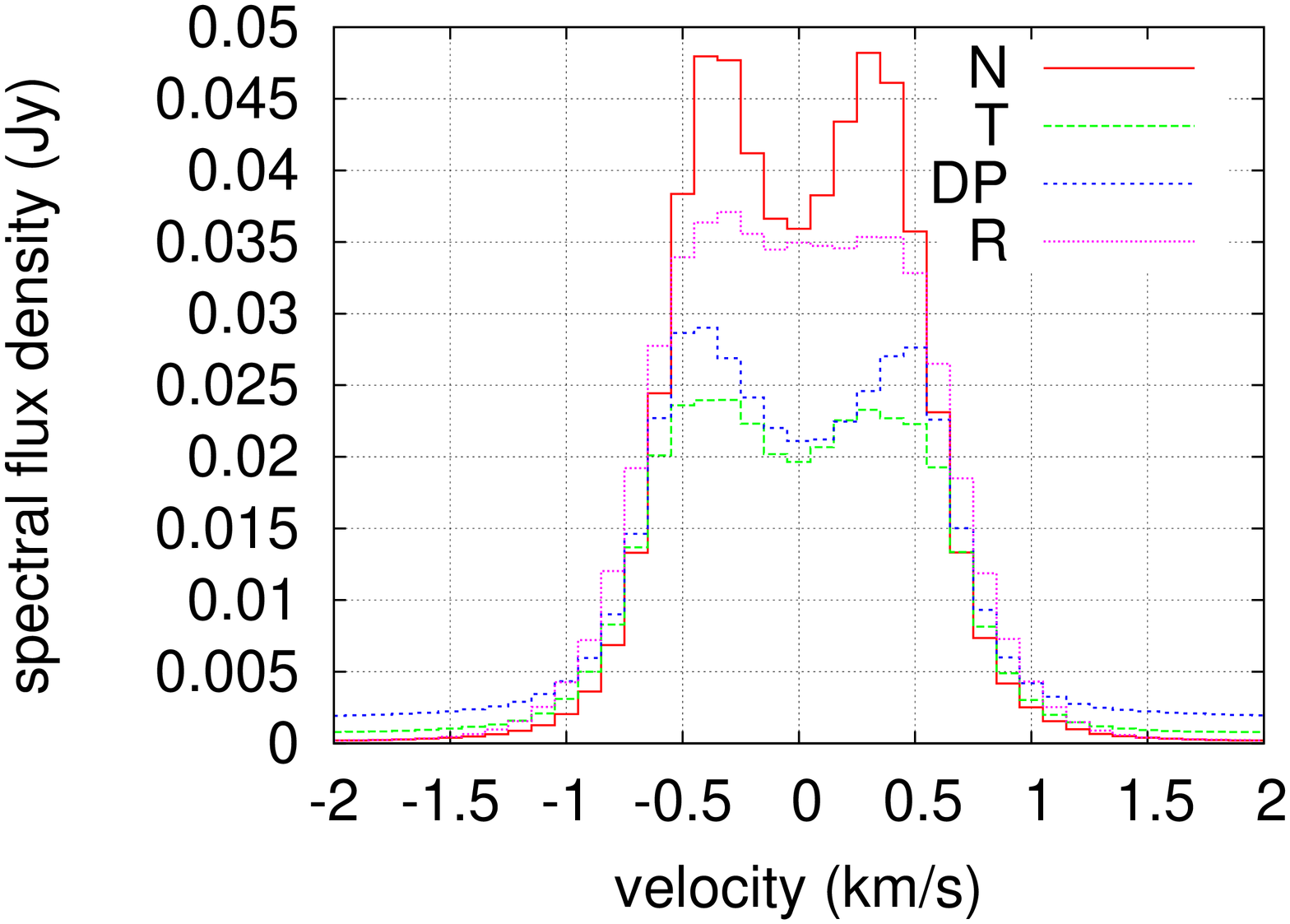}
   \caption{
    The line profiles for  $\mathrm{HCO}^{+}$ , $\mathrm{DCO}^{+}$ and $\mathrm{N_2H}^{+}$,
      assuming that the lightning takes place at $ \mathrm{50au} < r <  \mathrm{100au}$ of the disk.
    The labels {\tt no}, {\tt T}, {\tt DP}, and {\tt R} for the curves corresponds to
    no lightning, Townsend breakdown model, Druyversteyn-Penning breakdown model and
    runaway breakdown model, respectively.
   }\label{fig-lightning-lp}
   \end{figure}

\begin{figure}
   \begin{tabular}{ccccccc}
   \includegraphics[angle=270,width=4.5cm]{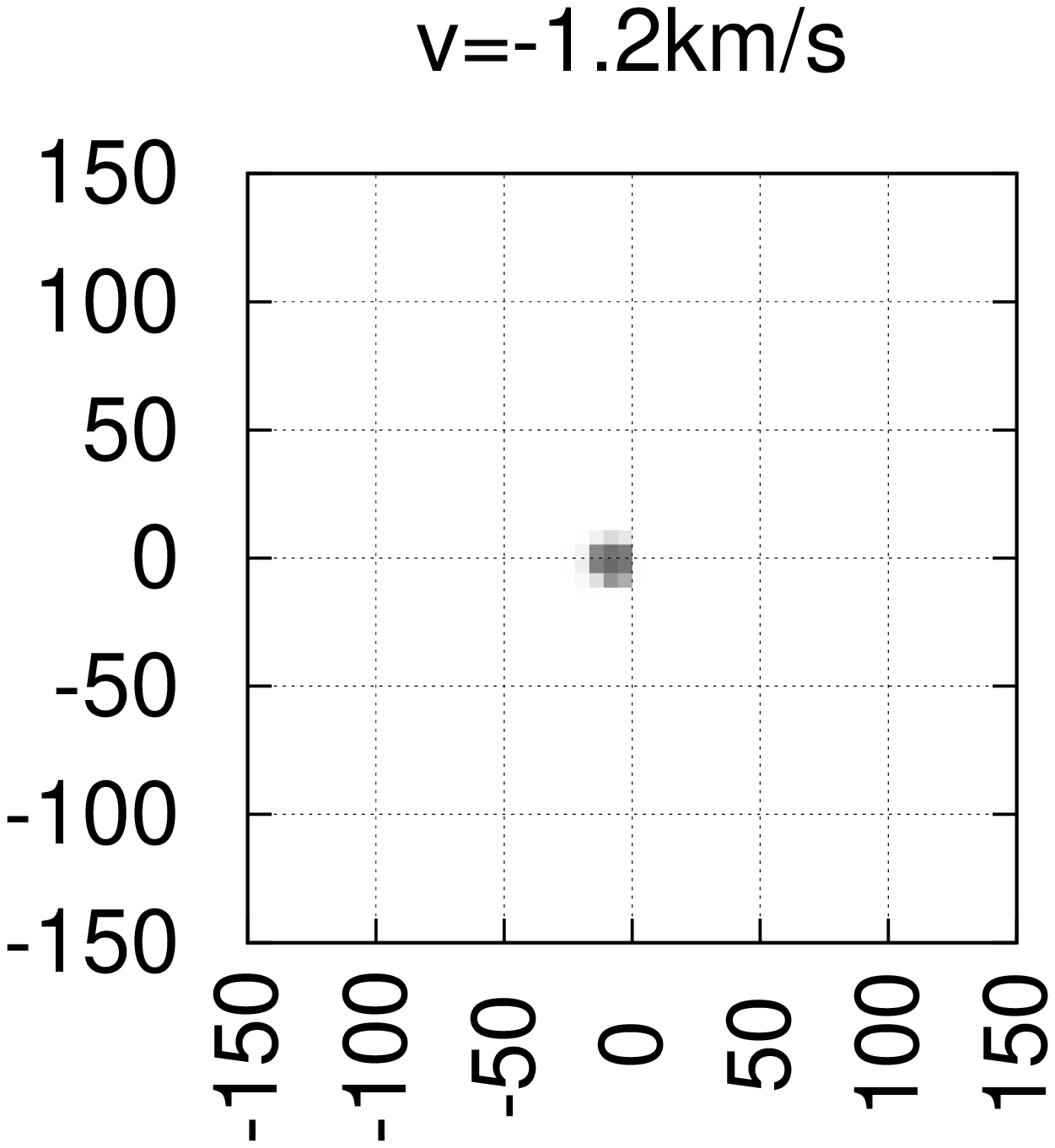}  \hspace{ -3.1cm}& \hspace{ -3.1cm}
   \includegraphics[angle=270,width=4.5cm]{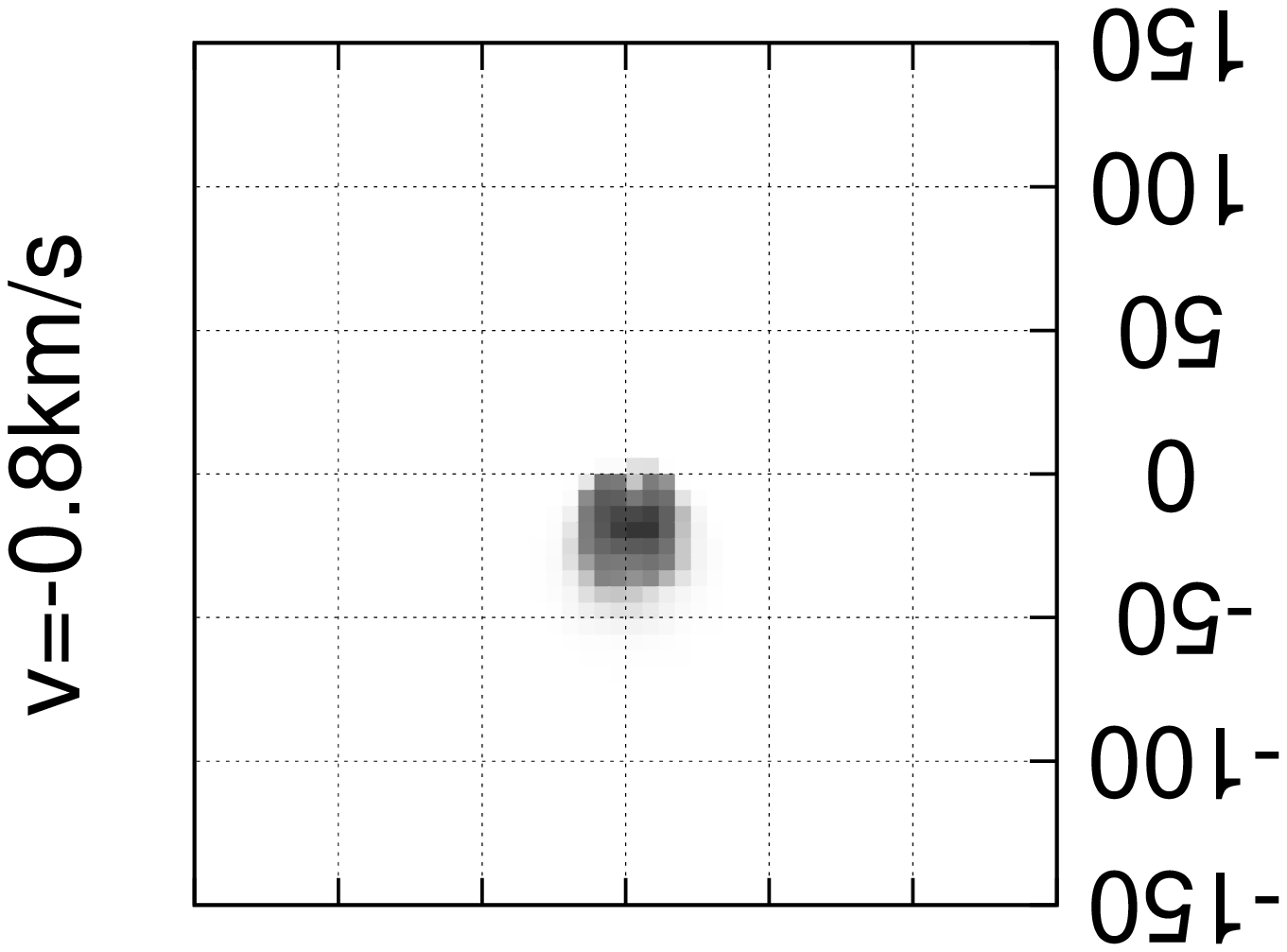}  \hspace{ -3.1cm}& \hspace{ -3.1cm}
   \includegraphics[angle=270,width=4.5cm]{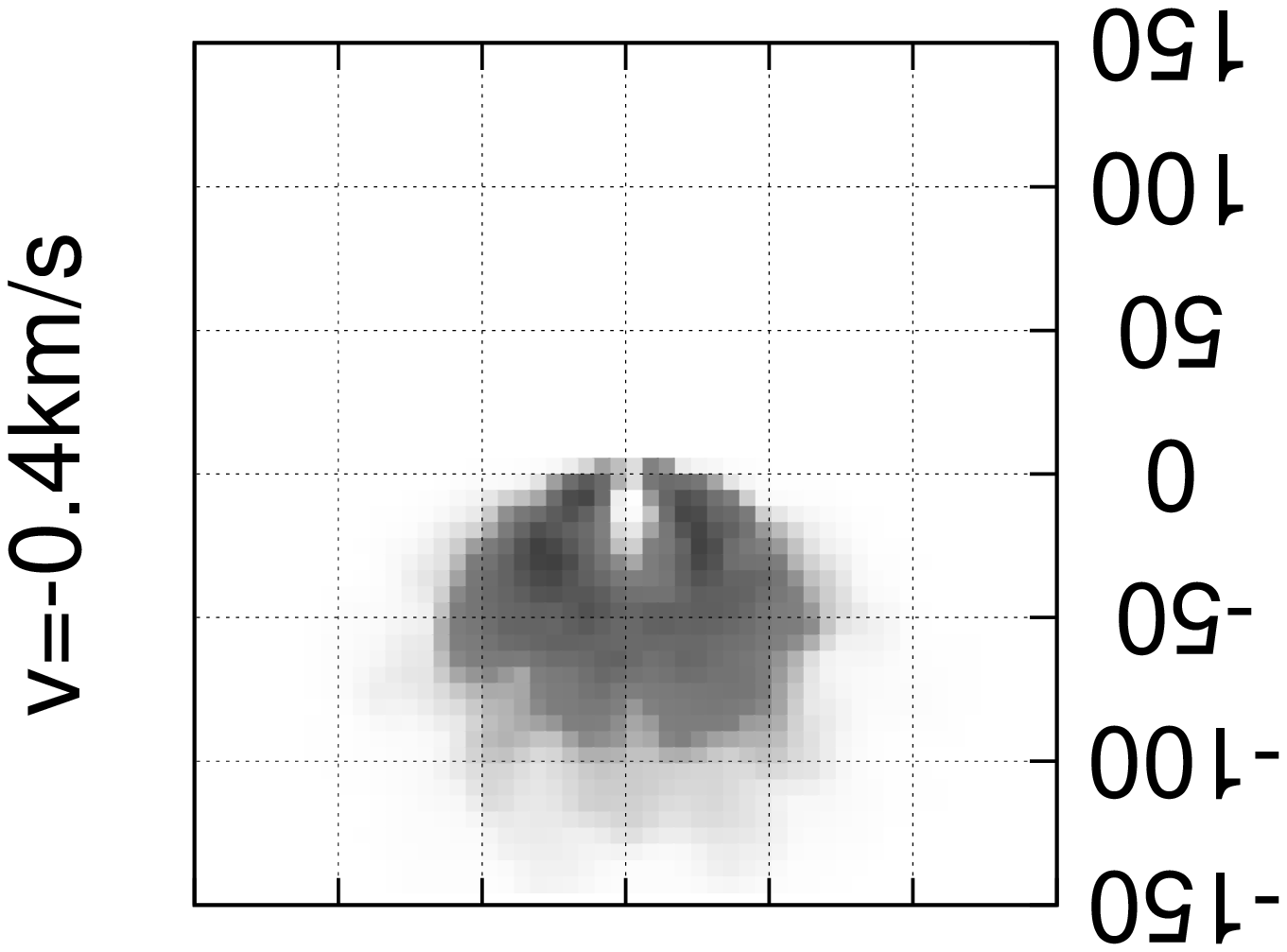}  \hspace{ -3.1cm}& \hspace{ -3.1cm}
   \includegraphics[angle=270,width=4.5cm]{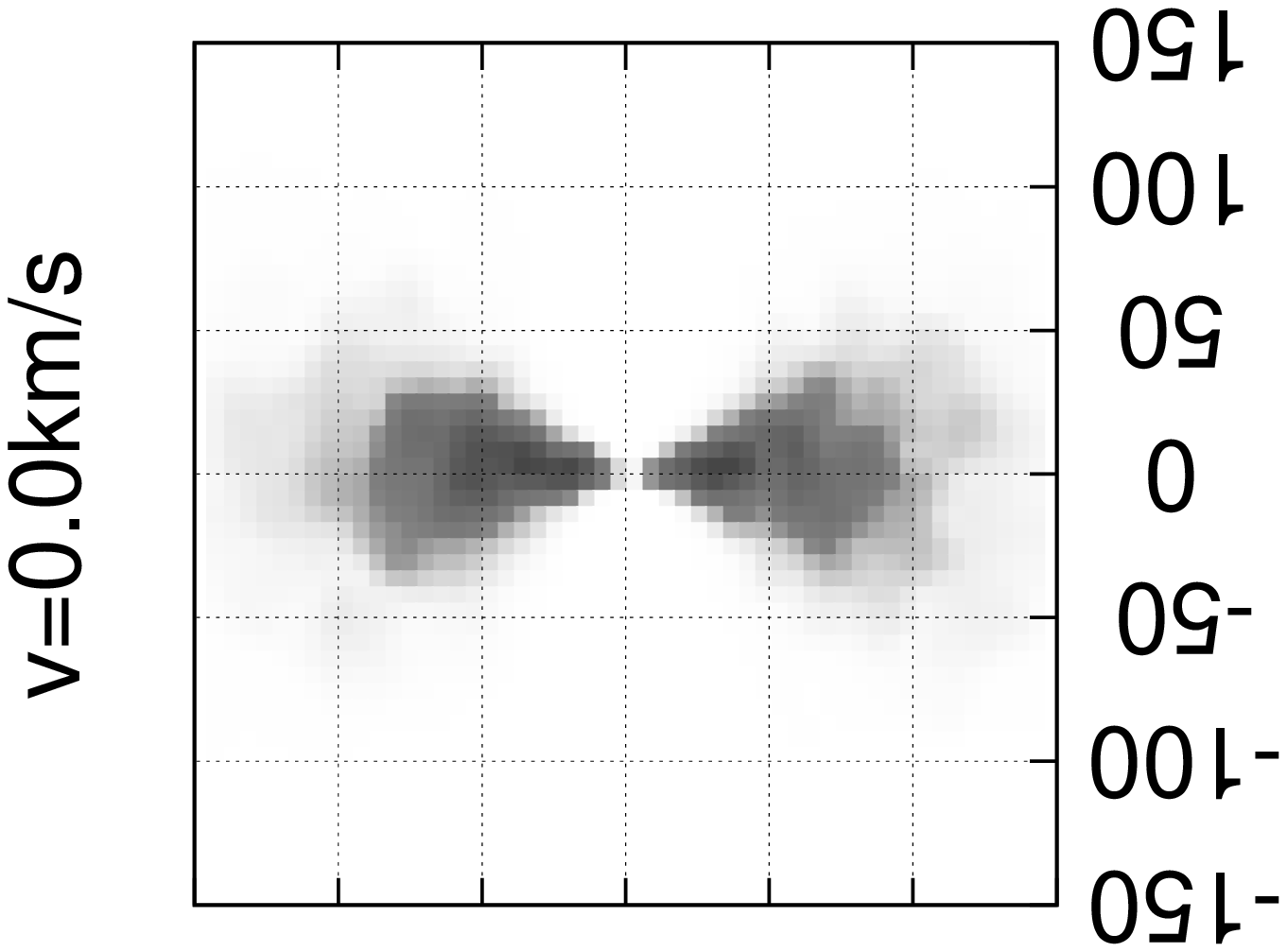}  \hspace{ -3.1cm}& \hspace{ -3.1cm}
   \includegraphics[angle=270,width=4.5cm]{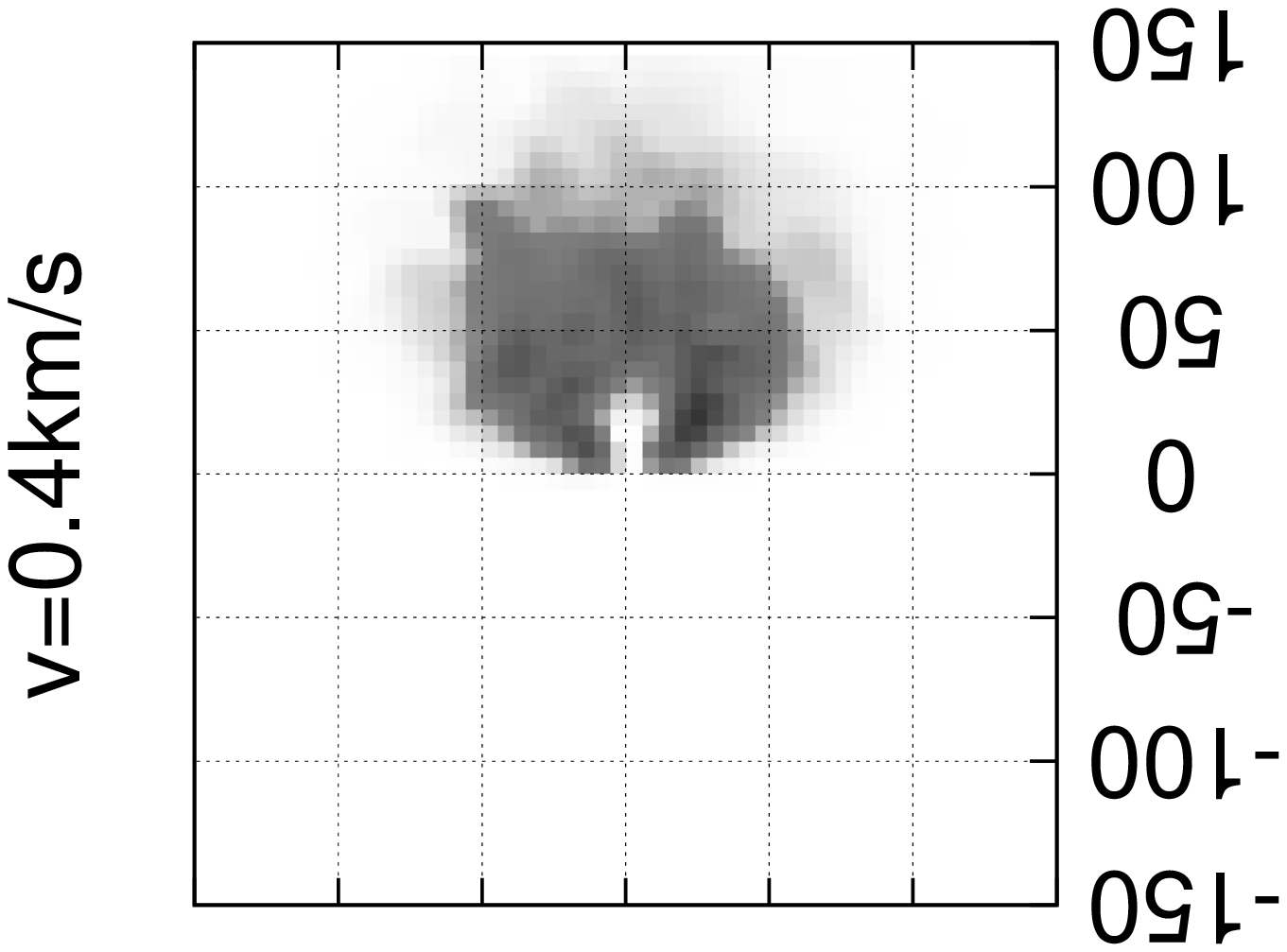}  \hspace{ -3.1cm}& \hspace{ -3.1cm}
   \includegraphics[angle=270,width=4.5cm]{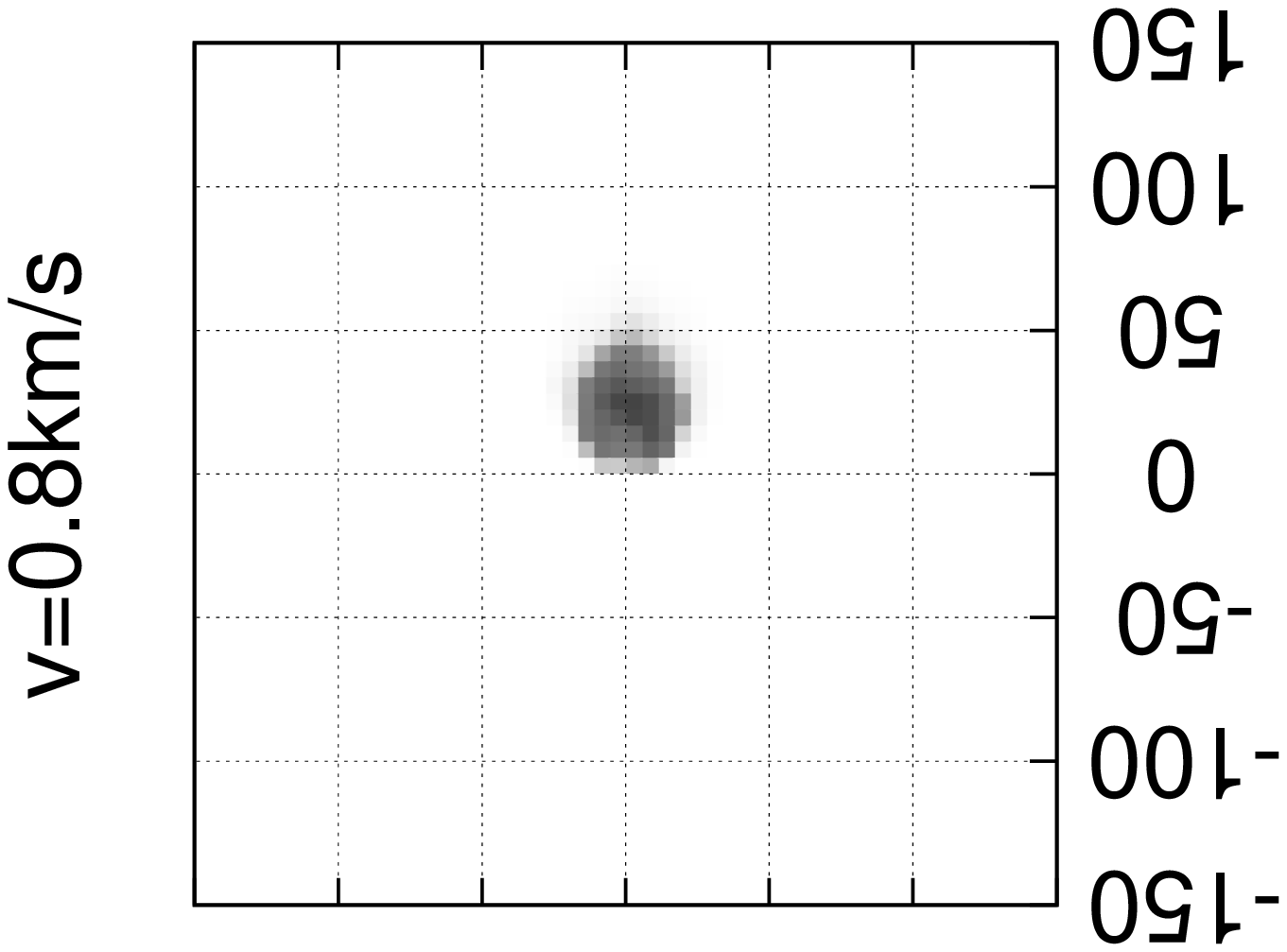}  \hspace{ -3.1cm}& \hspace{ -3.1cm}
   \includegraphics[angle=270,width=4.5cm]{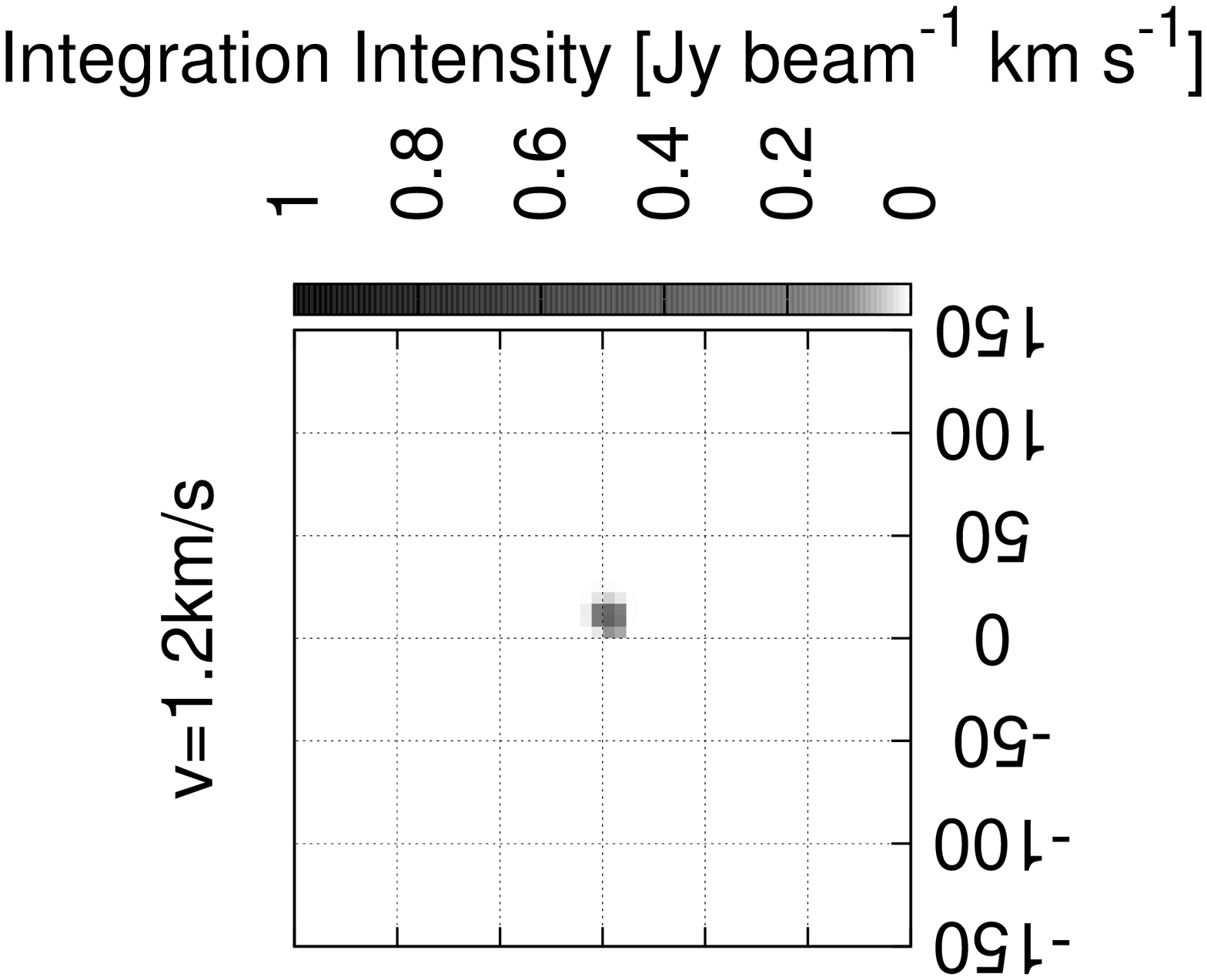}  \\
   \includegraphics[angle=270,width=4.5cm]{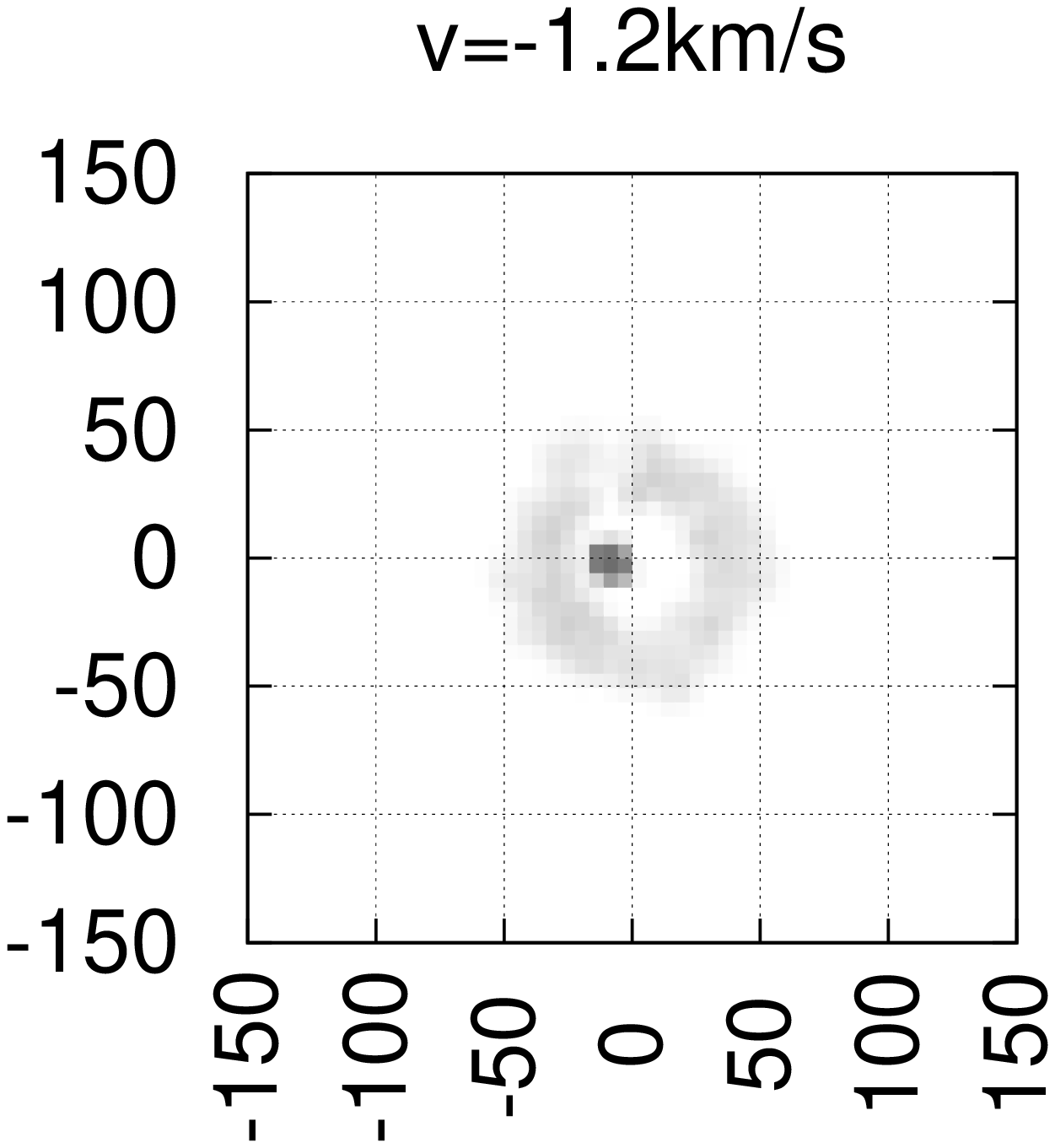}  \hspace{ -3.1cm}& \hspace{ -3.1cm}
   \includegraphics[angle=270,width=4.5cm]{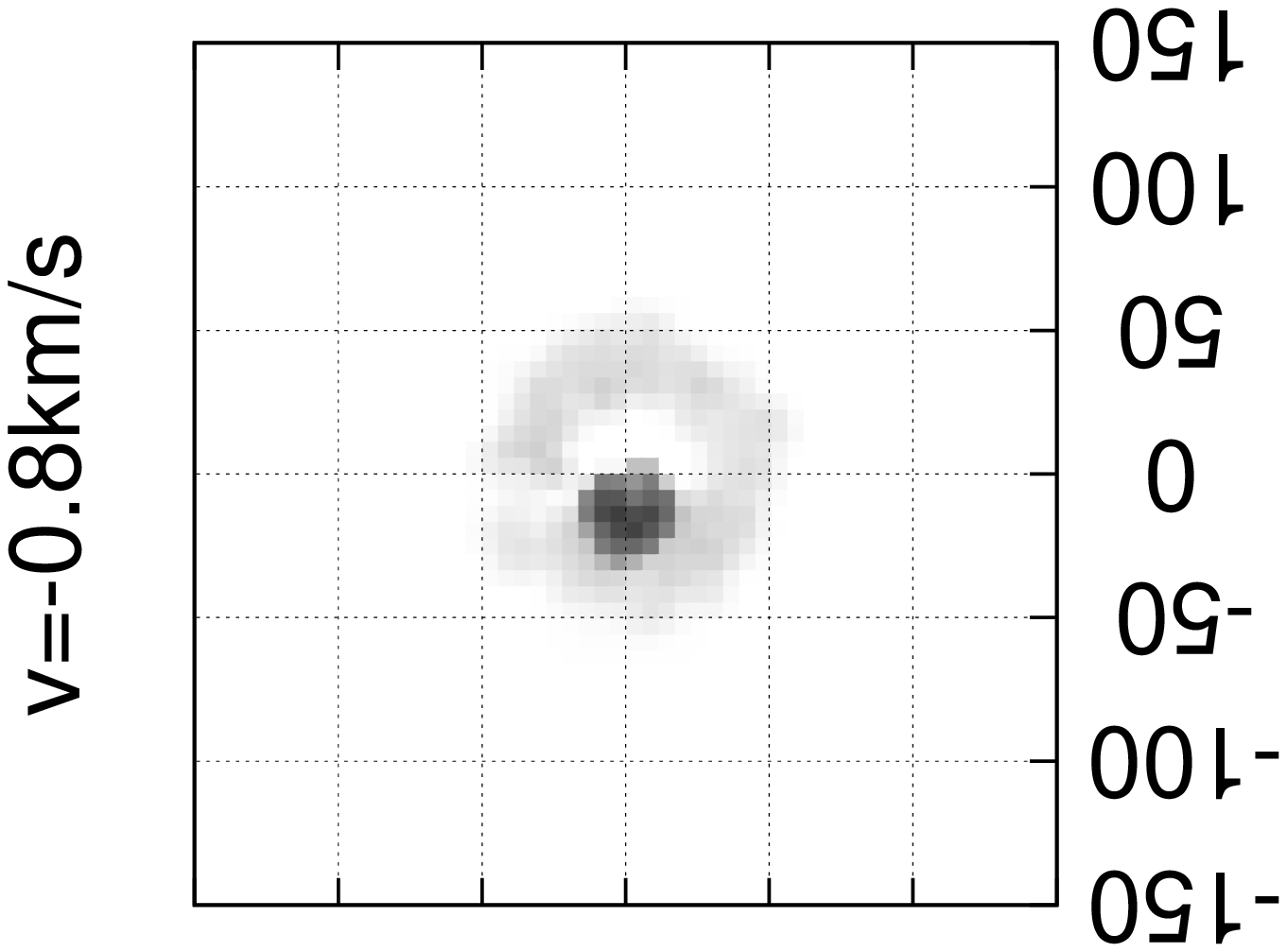}  \hspace{ -3.1cm}& \hspace{ -3.1cm}
   \includegraphics[angle=270,width=4.5cm]{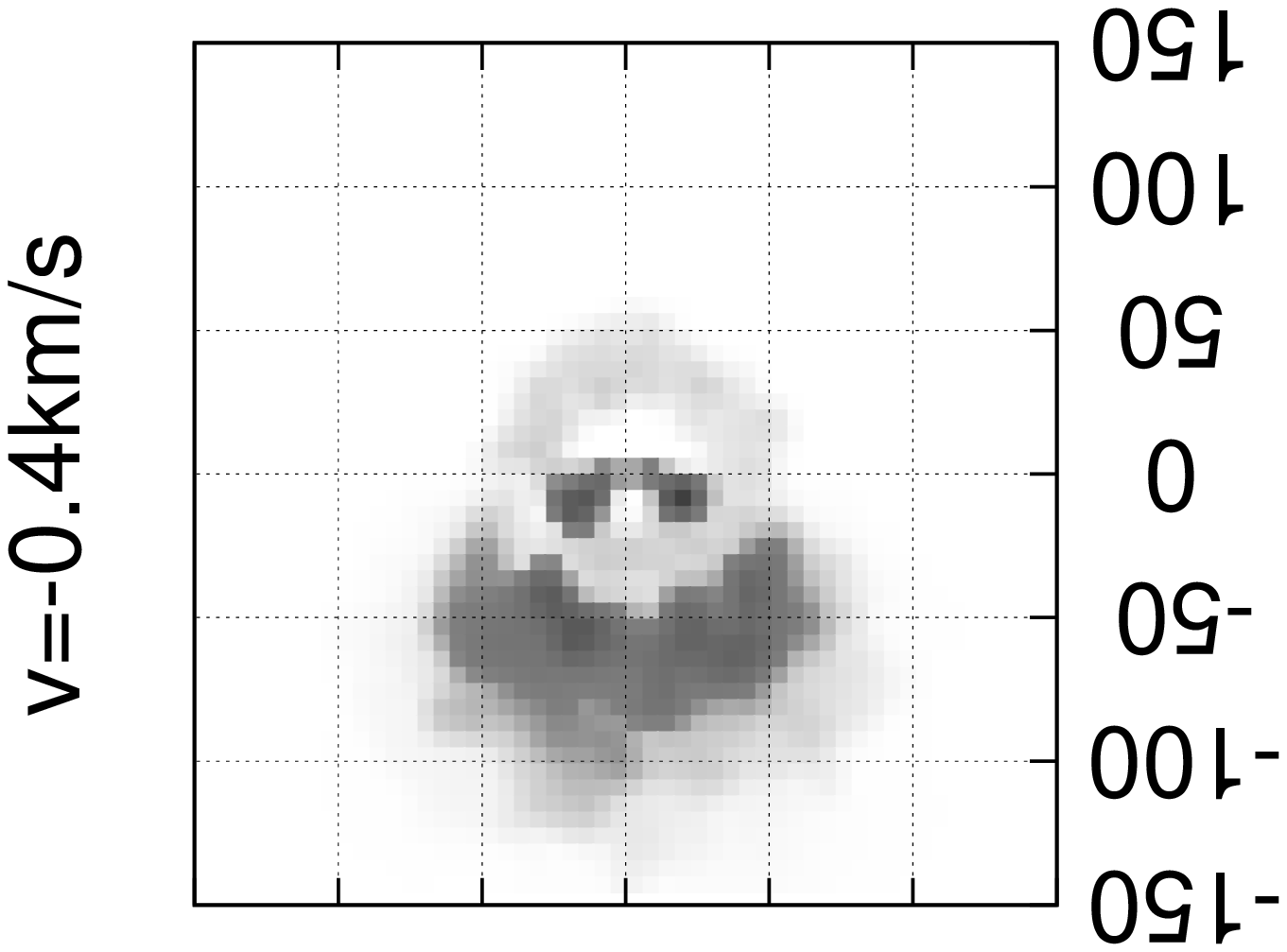}  \hspace{ -3.1cm}& \hspace{ -3.1cm}
   \includegraphics[angle=270,width=4.5cm]{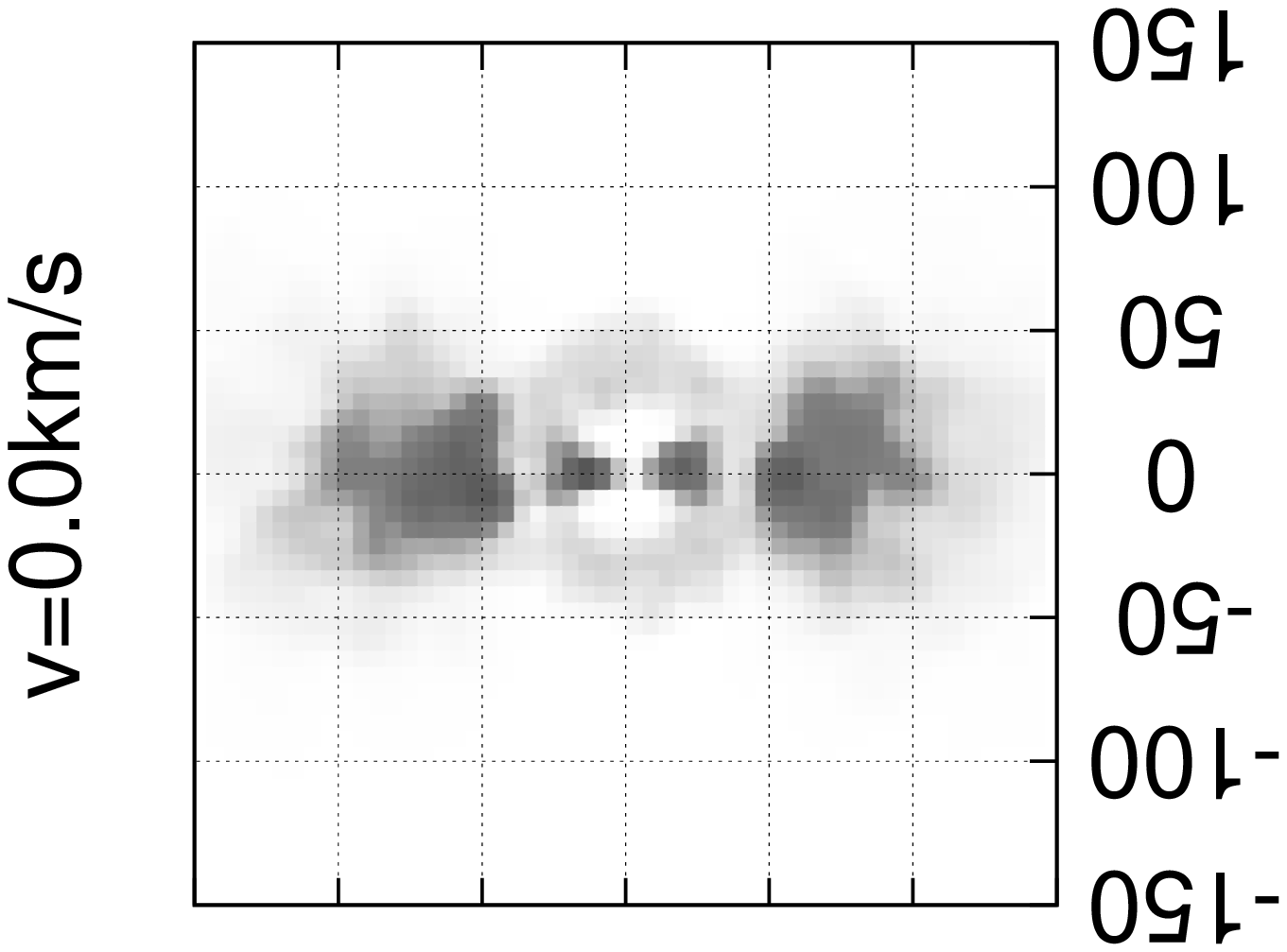}  \hspace{ -3.1cm}& \hspace{ -3.1cm}
   \includegraphics[angle=270,width=4.5cm]{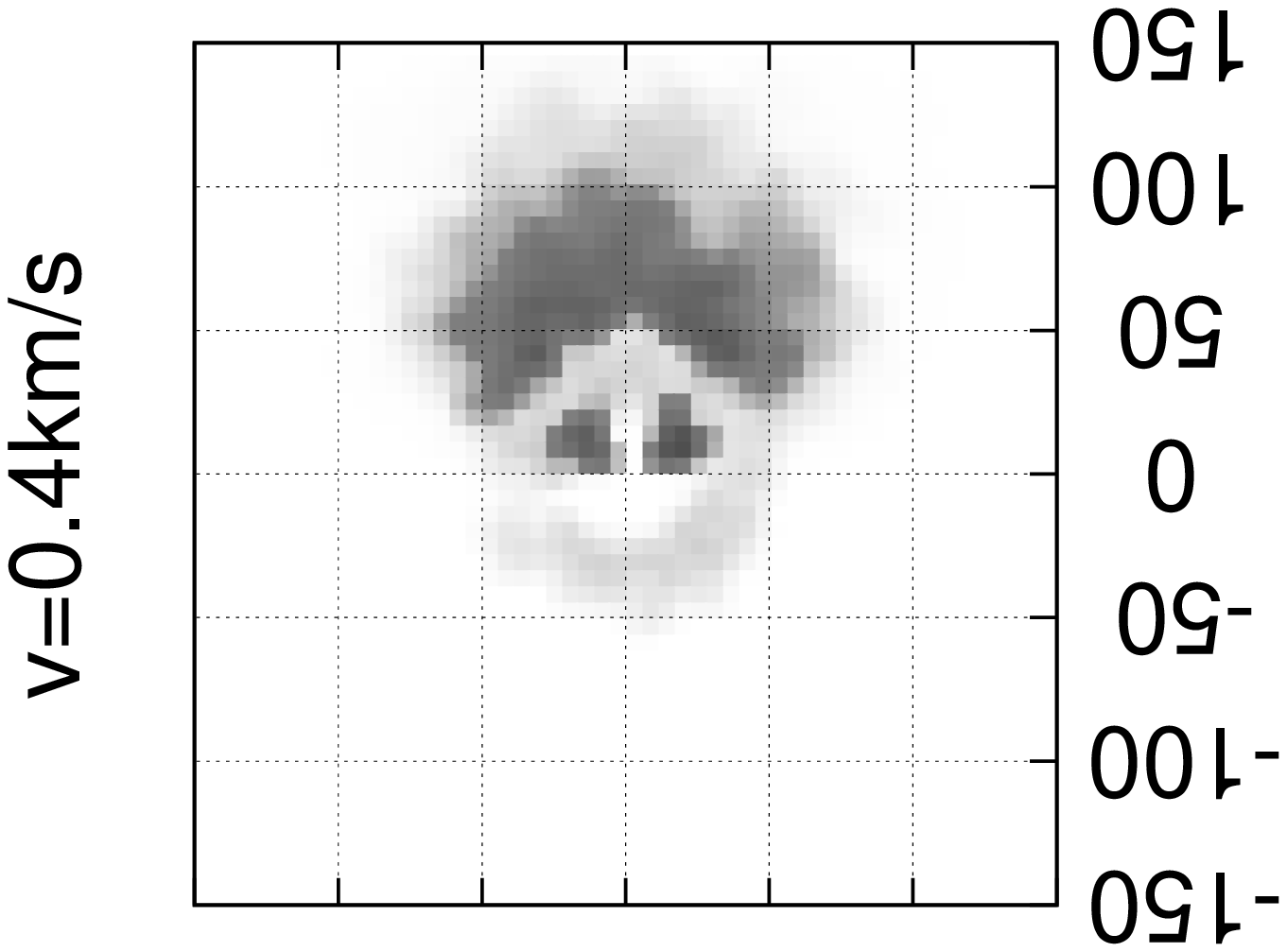}  \hspace{ -3.1cm}& \hspace{ -3.1cm}
   \includegraphics[angle=270,width=4.5cm]{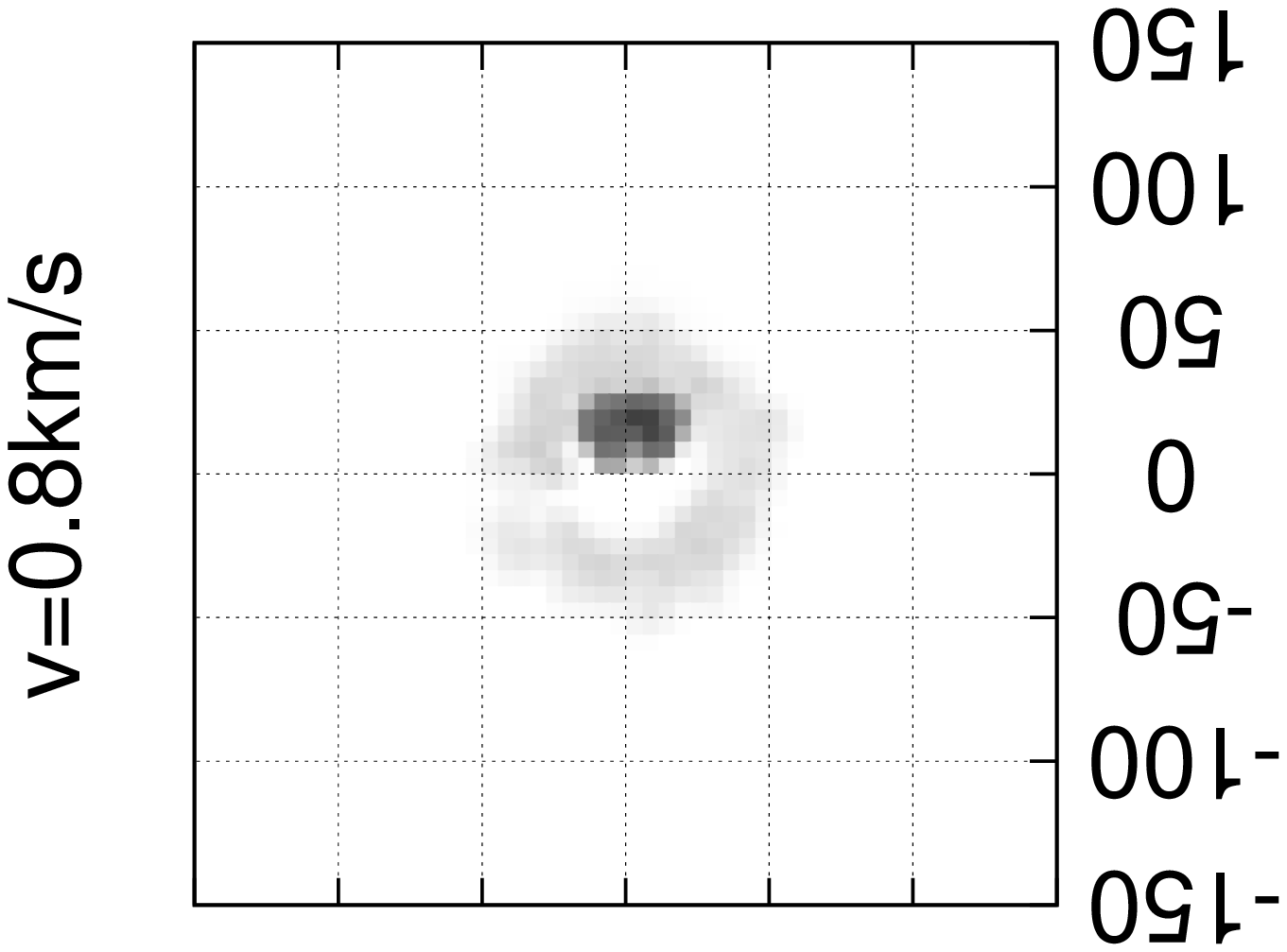}  \hspace{ -3.1cm}& \hspace{ -3.1cm}
   \includegraphics[angle=270,width=4.5cm]{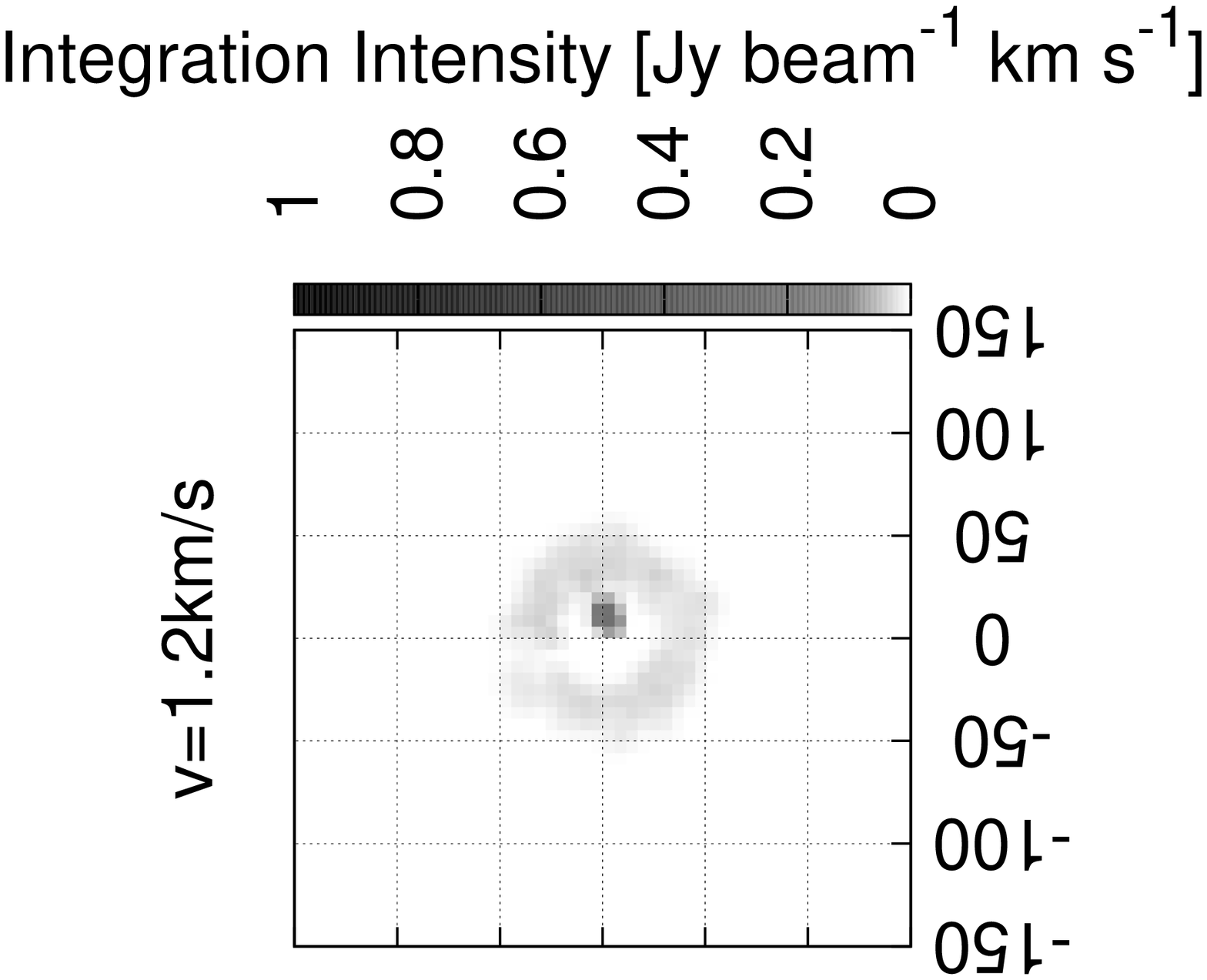}  \\
   \includegraphics[angle=270,width=4.5cm]{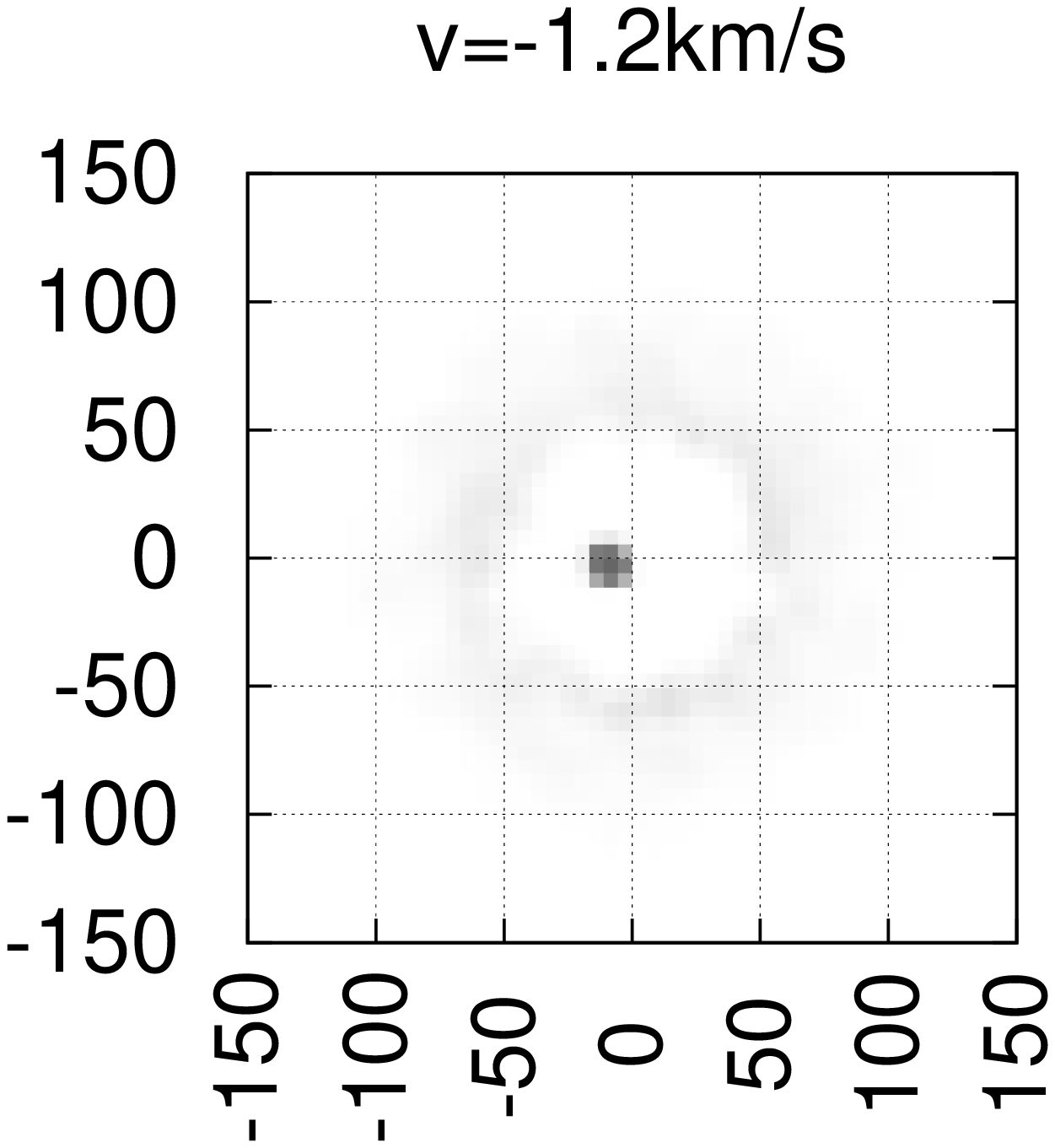}  \hspace{ -3.1cm}& \hspace{ -3.1cm}
   \includegraphics[angle=270,width=4.5cm]{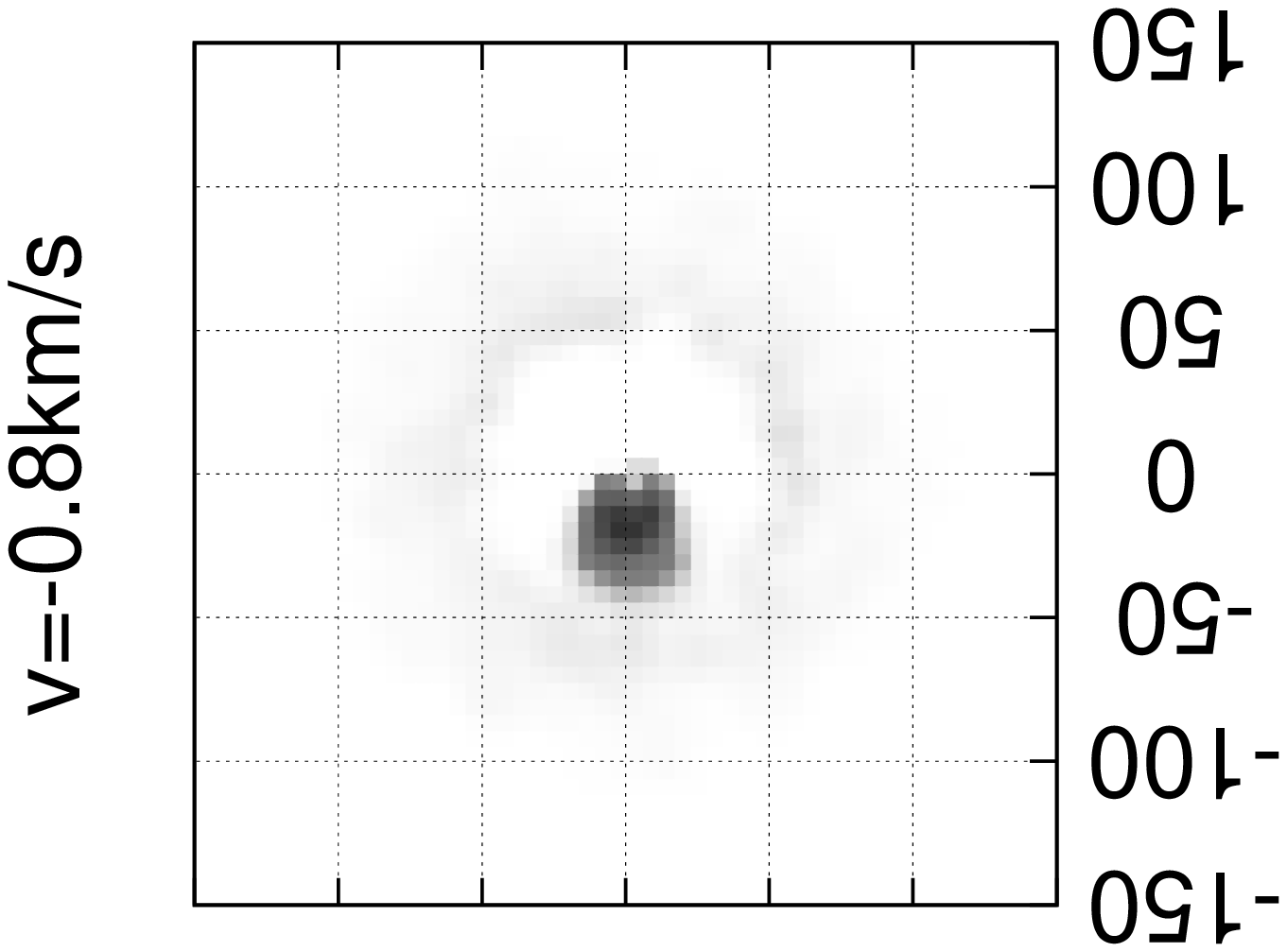}  \hspace{ -3.1cm}& \hspace{ -3.1cm}
   \includegraphics[angle=270,width=4.5cm]{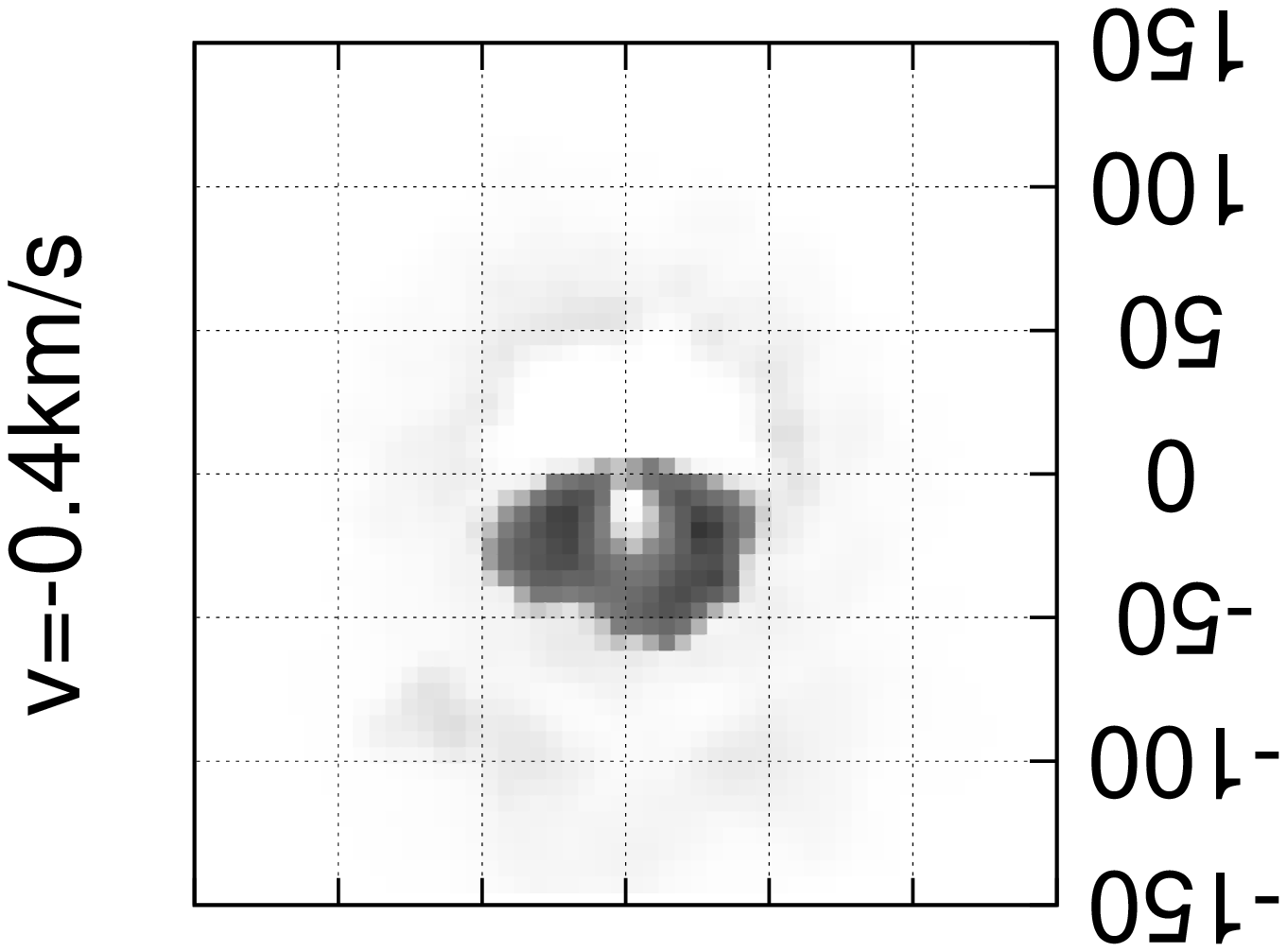}  \hspace{ -3.1cm}& \hspace{ -3.1cm}
   \includegraphics[angle=270,width=4.5cm]{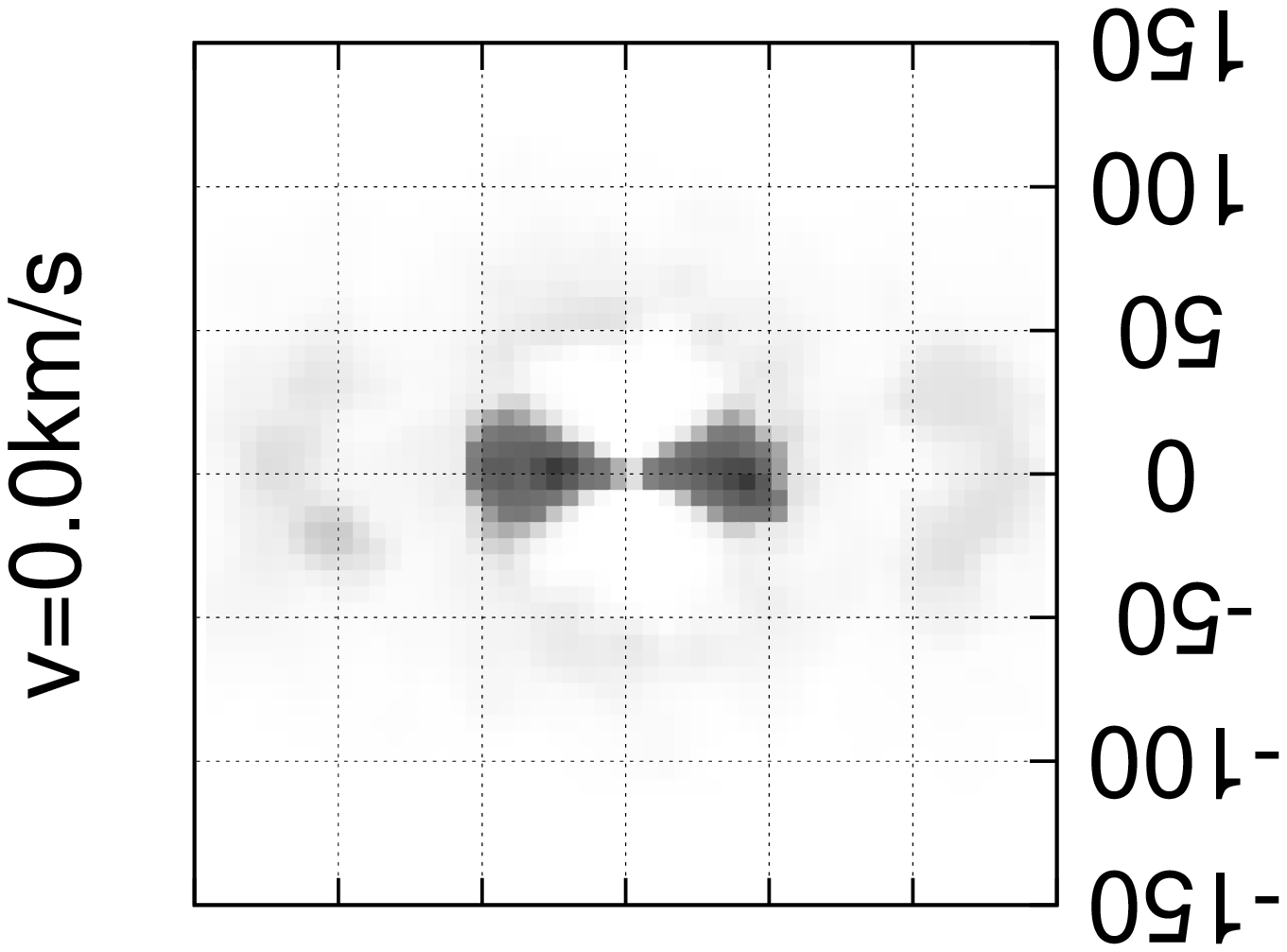}  \hspace{ -3.1cm}& \hspace{ -3.1cm}
   \includegraphics[angle=270,width=4.5cm]{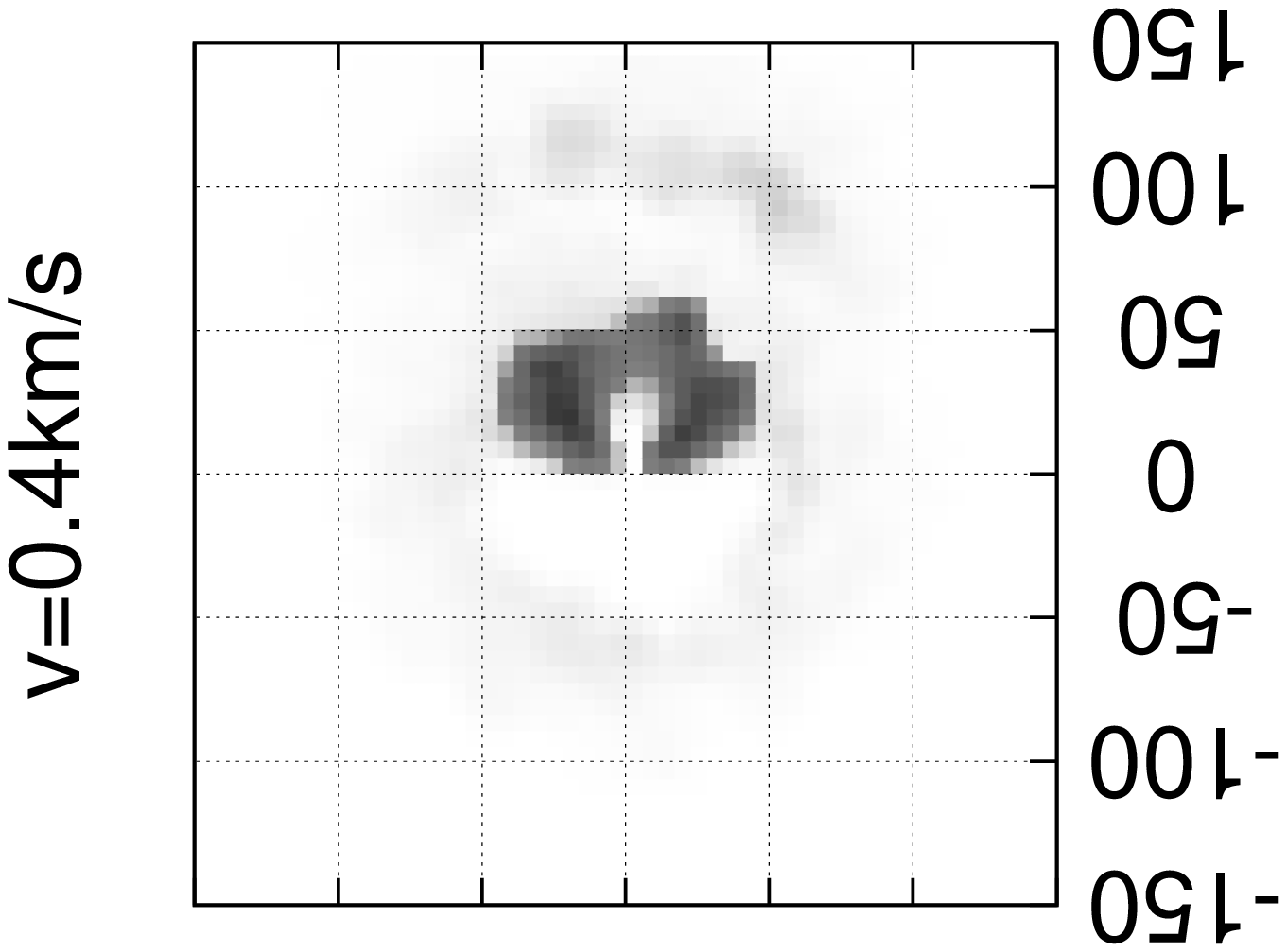}  \hspace{ -3.1cm}& \hspace{ -3.1cm}
   \includegraphics[angle=270,width=4.5cm]{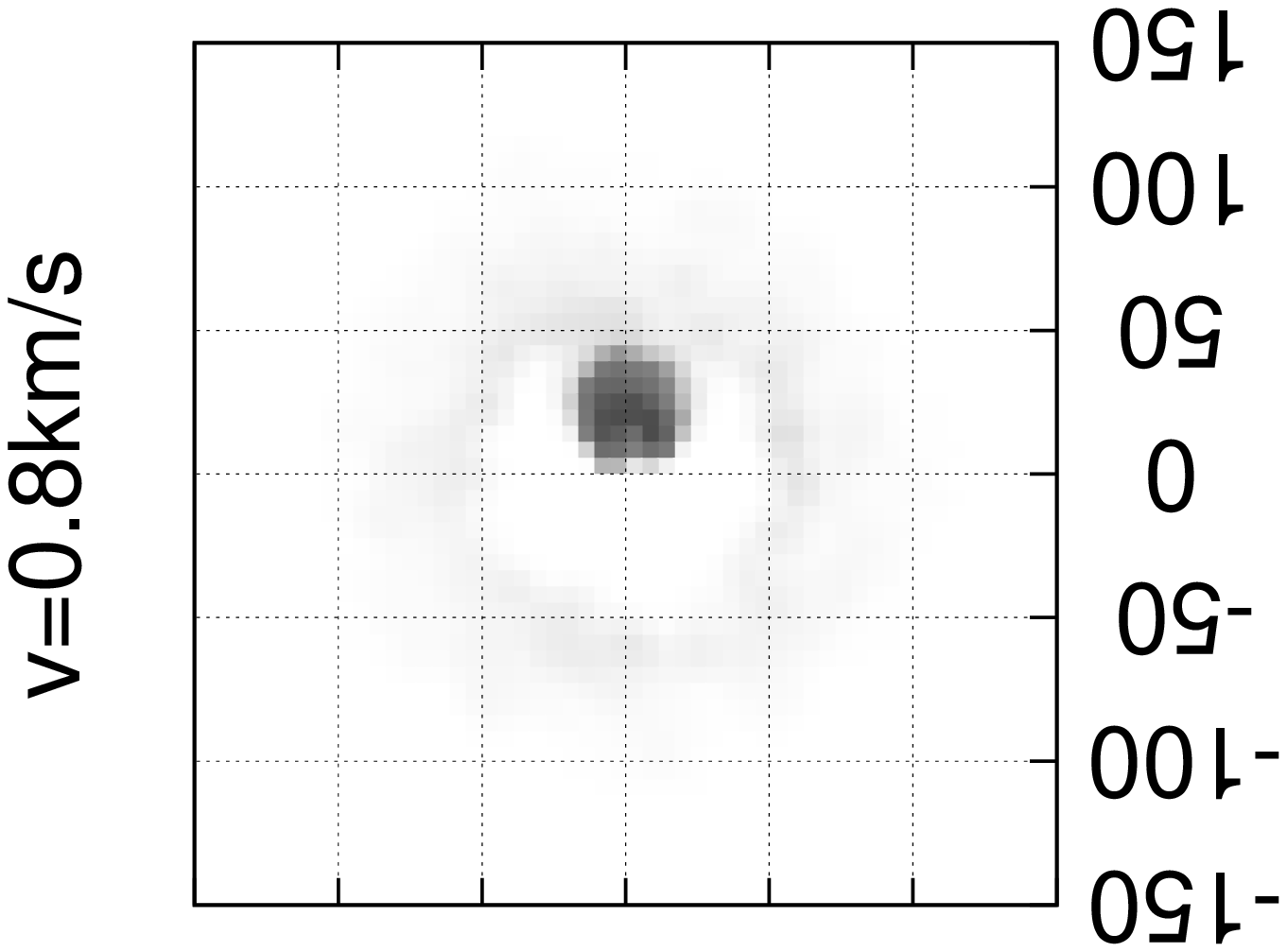}  \hspace{ -3.1cm}& \hspace{ -3.1cm}
   \includegraphics[angle=270,width=4.5cm]{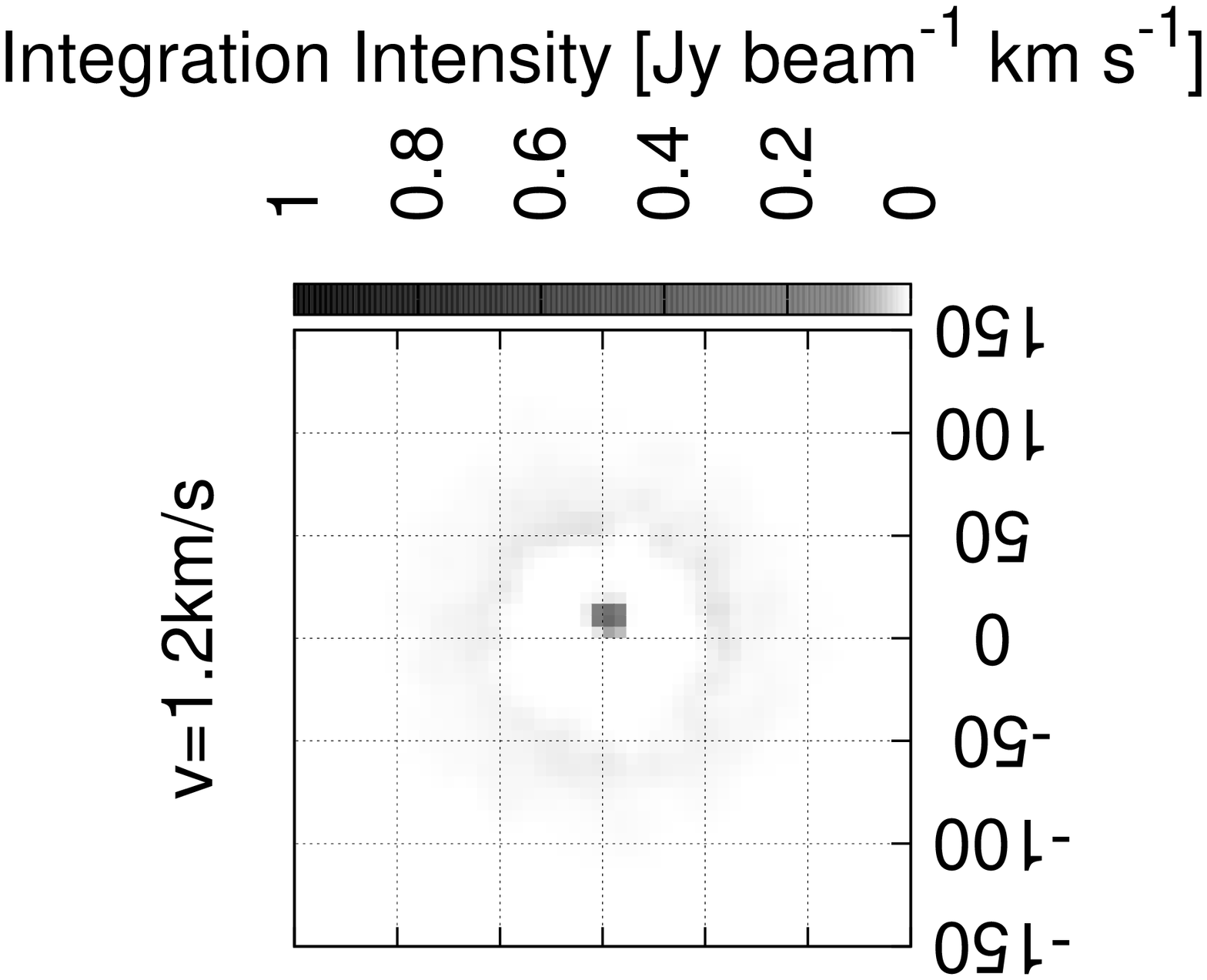}
   \end{tabular}
   \caption{
   Simulated {channel maps} of  $\mathrm{HCO}^{+}$ lines for
   disk models N (upper row),
   T25 (middle row), and
   T50  (lower row), respectively.
   Units are in $\mathrm{Jy} ~ \mathrm{beam}^{ -1} ~ \mathrm{km} ~\mathrm{s}^{ -1}$.
   We assume the beam size of $0''.65 \times 0''44$.
   \label{figEmissionMap}
   }
   \end{figure}

    We introduce seven disk models, as in Table \ref{tbl:DiskModels}.
    We calculate the line profiles for the three ion species with these seven disk models,
    in order to study the ability to distinguish the lightning model from the line observations (Figure \ref{fig-lightning-lp}).
    The line profiles are obtained by performing the spectral irradiance integral (equation (\ref{eq:SpectralIrradiance})).
    We assume isotropic distribution for the ion velocities, assuming that the electric field is turbulent.
    We simulate the {channel maps} using the
    spectral line radiation transfer code LIME by \citet{{doi:10.1051/0004-6361/201015333}}.
    In Figure \ref{figEmissionMap}, we present the
    simulated {channel maps} maps of the $\mathrm{HCO}^{+}$ line for
    N, T25, and T50 disk.
    We assumed that our model disk is located in the same way as
    TW Hya. That is, our model disk at the distance of $54~{\rm pc}$ and the
    inclination angle of $7^\circ$ \citep{{bibcode:2007ASSL..350.....V}}. Although we limit the disk parameters to this specific distance and inclination thoroughout this paper,
    our programs can be easily applied to other disk parameters.

  \subsection{Matched Filtering} \label{sec:MatchedFilter}

   We apply the matched filtering method \citep{{doi:10.1109/PROC.1963.2383}},
   in order to distinguish lightning model by ALMA.
   Matched filtering is
   the optimal method for discriminating models under noisy observation has been well studied
   and have wide  applications not only in radio astronomy
   \citep{{bibcode:2003ApJS..147..167E}} but also in
   extra solar planet astronomy \citep{{bibcode:1996Icar..119..244J,bibcode:2000ApJ...535..338D}},
   gravitational wave astronomy \citep{{bibcode:1999PhRvD..60b2002O,bibcode:2007PhRvD..76h4020V,bibcode:2013PhRvD..88d4026H}},
   and even in ocean tomography \citep{{bibcode:1979DSRA...26..123M}}.
   We follow the treatment by \citet{{isbn:978-3-527-40886-3}} .

   Given that the noise levels for $\mathrm{HCO}^{+}$,  $\mathrm{DCO}^{+}$ and  $\mathrm{N_2H}^{+}$ are
   $1.130 \times 10^{-2}~{\rm Jy}$,
   $1.330 \times 10^{-2}~{\rm Jy}$, and
   $1.800 \times 10^{-2}~{\rm Jy}$, respectively,
   their noise spectrum power density $S_h$ per square arcsecond are
   $5.843 \times 10^{-5}~{\rm Jy^2~km/s}$,
   $8.094 \times 10^{-5}~{\rm Jy^2~km/s}$, and
   $1.483 \times 10^{-4}~{\rm Jy^2~km/s}$, respectively.

   The measure-of-sensitivity $\sigma_{\rm mos}$ of the matched-filter
   between two images $h_1(x,y,v)$ and $h_2(x,y,v)$ is:
   \begin{eqnarray}
      \sigma_{\rm mos} = 4\int_{v_{\rm min}}^{v_{\rm max}} \int_{(x,y)\in {\rm image}}
      \frac{|h_1(x,y,v) - h_2(x,y,v)|^2}{S_h} \mathit{dx}  \mathit{dy} \mathit{dv} .
   \end{eqnarray}
   Here, $x$ and $y$ are image coordinates in arcseconds, and $v$ is the velocity coordinate.

  \begin{table}[hp]
  \begin{tabular}{cc}
  
  \begin{tabular}{|c|ccc|}
  \hline
  $\mathrm{HCO}^{+}$ &  T25 & DP25 & R25 \\
  \hline
  N      &  3729.7 &  3021.6 &  1277.1 \\
  T25  &               & 1508.4 &  3294.1 \\
  DP25 &               &                &  2608.7 \\
  \hline
  \multicolumn{1}{c}{\phantom{888 species}} &
  \multicolumn{1}{c}{\phantom{888.88}} &
  \multicolumn{1}{c}{\phantom{888.88}} &
  \multicolumn{1}{c}{\phantom{888.88}}
  \end{tabular}
    & 
  \begin{tabular}{|c|ccc|}
  \hline
  $\mathrm{HCO}^{+}$ &  T50 & DP50 & R50 \\
  \hline
  N      &  2488.4 &  2199.2 &  1277.4 \\
  T50  &               & 1418.7 &  2316.5 \\
  DP50 &               &                &  2125.5 \\
  \hline
  \multicolumn{1}{c}{\phantom{888 species}} &
  \multicolumn{1}{c}{\phantom{888.88}} &
  \multicolumn{1}{c}{\phantom{888.88}} &
  \multicolumn{1}{c}{\phantom{888.88}}
  \end{tabular}
    \\
  
  \begin{tabular}{|c|ccc|}
  \hline
  $\mathrm{DCO}^{+}$ &  T25 & DP25 & R25 \\
  \hline
  N      &  122.3 &  104.9 &  46.3 \\
  T25  &               & 44.7 &  115.3 \\
  DP25 &               &                &  99.3 \\
  \hline
  \multicolumn{1}{c}{\phantom{888 species}} &
  \multicolumn{1}{c}{\phantom{888.88}} &
  \multicolumn{1}{c}{\phantom{888.88}} &
  \multicolumn{1}{c}{\phantom{888.88}}
  \end{tabular}
    & 
  \begin{tabular}{|c|ccc|}
  \hline
  $\mathrm{DCO}^{+}$ &  T50 & DP50 & R50 \\
  \hline
  N      &  111.3 &  100.9 &  45.8 \\
  T50  &               & 42.8 &  95.3 \\
  DP50 &               &                &  81.6 \\
  \hline
  \multicolumn{1}{c}{\phantom{888 species}} &
  \multicolumn{1}{c}{\phantom{888.88}} &
  \multicolumn{1}{c}{\phantom{888.88}} &
  \multicolumn{1}{c}{\phantom{888.88}}
  \end{tabular}
    \\
  
  \begin{tabular}{|c|ccc|}
  \hline
  $\mathrm{N_2H}^{+}$ &  T25 & DP25 & R25 \\
  \hline
  N      &  5.8 &  5.2 &  2.3 \\
  T25  &               & 2.0 &  5.8 \\
  DP25 &               &                &  5.1 \\
  \hline
  \multicolumn{1}{c}{\phantom{888 species}} &
  \multicolumn{1}{c}{\phantom{888.88}} &
  \multicolumn{1}{c}{\phantom{888.88}} &
  \multicolumn{1}{c}{\phantom{888.88}}
  \end{tabular}
    & 
  \begin{tabular}{|c|ccc|}
  \hline
  $\mathrm{N_2H}^{+}$ &  T50 & DP50 & R50 \\
  \hline
  N      &  4.0 &  3.5 &  2.4 \\
  T50  &               & 2.3 &  3.8 \\
  DP50 &               &                &  3.4 \\
  \hline
  \multicolumn{1}{c}{\phantom{888 species}} &
  \multicolumn{1}{c}{\phantom{888.88}} &
  \multicolumn{1}{c}{\phantom{888.88}} &
  \multicolumn{1}{c}{\phantom{888.88}}
  \end{tabular}
    \\
  
  \begin{tabular}{|c|ccc|}
  \hline
  3 species &  T25 & DP25 & R25 \\
  \hline
  N      &  3857.9 &  3131.7 &  1325.6 \\
  T25  &               & 1555.2 &  3415.1 \\
  DP25 &               &                &  2713.0 \\
  \hline
  \multicolumn{1}{c}{\phantom{888 species}} &
  \multicolumn{1}{c}{\phantom{888.88}} &
  \multicolumn{1}{c}{\phantom{888.88}} &
  \multicolumn{1}{c}{\phantom{888.88}}
  \end{tabular}
    & 
  \begin{tabular}{|c|ccc|}
  \hline
  3 species &  T50 & DP50 & R50 \\
  \hline
  N      &  2603.7 &  2303.7 &  1325.6 \\
  T50  &               & 1463.9 &  2415.5 \\
  DP50 &               &                &  2210.4 \\
  \hline
  \multicolumn{1}{c}{\phantom{888 species}} &
  \multicolumn{1}{c}{\phantom{888.88}} &
  \multicolumn{1}{c}{\phantom{888.88}} &
  \multicolumn{1}{c}{\phantom{888.88}}
  \end{tabular}
   \\
  \end{tabular}
  \caption{
  The measure of sensitivity values among {\tt N}, {\tt T25}, {\tt DP25} and {\tt R25} models,
  and among {\tt N}, {\tt T50}, {\tt DP50} and {\tt R50} models,
  using either one of,
  or all the three of, our lines
  $\mathrm{HCO}^{+}~3-2$ ,
  $\mathrm{DCO}^{+}~3-2$ and $\mathrm{N_2H}^{+}~3-2$.
  }\label{tbl:MOS}
  \end{table}

  \begin{table}[hp]
  \begin{tabular}{cc}
   \multicolumn{2}{c}{XL} \\
  
  \begin{tabular}{|c|ccc|}
  \hline
  3 species &  T25 & DP25 & R25 \\
  \hline
  N      &  2872.2 &  1547.5 &  1193.0 \\
  T25  &               & 1988.4 &  2813.6 \\
  DP25 &               &                &  1543.8 \\
  \hline
  \multicolumn{1}{c}{\phantom{888 species}} &
  \multicolumn{1}{c}{\phantom{888.88}} &
  \multicolumn{1}{c}{\phantom{888.88}} &
  \multicolumn{1}{c}{\phantom{888.88}}
  \end{tabular}
    & 
  \begin{tabular}{|c|ccc|}
  \hline
  3 species &  T50 & DP50 & R50 \\
  \hline
  N      &  2365.7 &  1539.3 &  1326.4 \\
  T50  &               & 1539.7 &  2408.2 \\
  DP50 &               &                &  1635.3 \\
  \hline
  \multicolumn{1}{c}{\phantom{888 species}} &
  \multicolumn{1}{c}{\phantom{888.88}} &
  \multicolumn{1}{c}{\phantom{888.88}} &
  \multicolumn{1}{c}{\phantom{888.88}}
  \end{tabular}
   \\
   \multicolumn{2}{c}{XM} \\
  
  \begin{tabular}{|c|ccc|}
  \hline
  3 species &  T25 & DP25 & R25 \\
  \hline
  N      &  3857.9 &  3131.7 &  1325.6 \\
  T25  &               & 1555.2 &  3415.1 \\
  DP25 &               &                &  2713.0 \\
  \hline
  \multicolumn{1}{c}{\phantom{888 species}} &
  \multicolumn{1}{c}{\phantom{888.88}} &
  \multicolumn{1}{c}{\phantom{888.88}} &
  \multicolumn{1}{c}{\phantom{888.88}}
  \end{tabular}
    & 
  \begin{tabular}{|c|ccc|}
  \hline
  3 species &  T50 & DP50 & R50 \\
  \hline
  N      &  2603.7 &  2303.7 &  1325.6 \\
  T50  &               & 1463.9 &  2415.5 \\
  DP50 &               &                &  2210.4 \\
  \hline
  \multicolumn{1}{c}{\phantom{888 species}} &
  \multicolumn{1}{c}{\phantom{888.88}} &
  \multicolumn{1}{c}{\phantom{888.88}} &
  \multicolumn{1}{c}{\phantom{888.88}}
  \end{tabular}
   \\
   \multicolumn{2}{c}{XS} \\
  
  \begin{tabular}{|c|ccc|}
  \hline
  3 species &  T25 & DP25 & R25 \\
  \hline
  N      &  3483.7 &  3484.1 &  2101.3 \\
  T25  &               & 1294.1 &  2315.4 \\
  DP25 &               &                &  2217.8 \\
  \hline
  \multicolumn{1}{c}{\phantom{888 species}} &
  \multicolumn{1}{c}{\phantom{888.88}} &
  \multicolumn{1}{c}{\phantom{888.88}} &
  \multicolumn{1}{c}{\phantom{888.88}}
  \end{tabular}
    & 
  \begin{tabular}{|c|ccc|}
  \hline
  3 species &  T50 & DP50 & R50 \\
  \hline
  N      &  2648.4 &  2623.6 &  2067.3 \\
  T50  &               & 1170.2 &  1683.7 \\
  DP50 &               &                &  1603.3 \\
  \hline
  \multicolumn{1}{c}{\phantom{888 species}} &
  \multicolumn{1}{c}{\phantom{888.88}} &
  \multicolumn{1}{c}{\phantom{888.88}} &
  \multicolumn{1}{c}{\phantom{888.88}}
  \end{tabular}
   \\
  \end{tabular}
  \caption{
    {
    The dependence of the measure of sensitivity on the cross section models.
  The measure of sensitivity was estimated among {\tt N}, {\tt T25}, {\tt DP25} and {\tt R25} models,
  and among {\tt N}, {\tt T50}, {\tt DP50} and {\tt R50} models,
  using all the three of, our lines
  $\mathrm{HCO}^{+}~3-2$ ,
  $\mathrm{DCO}^{+}~3-2$ and $\mathrm{N_2H}^{+}~3-2$.
  }}\label{tbl:MOS2}
  \end{table}

   The measure of sensitivity among the models using different lines are summarized in
   Table \ref{tbl:MOS}.
   The measure-of-sensitivity for any two different models is larger than 100,
   and the largest measure-of-sensitivity is greater than 1000.
   Therefore the image like Figure \ref{figEmissionMap} is not difficult to detect.
   However, no observation of protoplanetary disk has been reported.
   Therefore, we can reject such form of lightning models from observations.
   There are multiple alternative scenarios that observations suggest:
   (1) Protoplanetary disk lightning does not exist at all. (2) The probability of protoplanetary disk
   with lightning matrix gas (LMG) is low, so that we have not yet observed one yet.
   (3) Protoplanetary disk LMG
   exists in forms of LMG clumps (protoplanetary ``cumulonibus clouds'') much smaller than the size of
   the protoplanetary disks,
   {(4) Panetary disk LMG is scattered in many smaller clumps in the protoplanetary disk with certain
   volume-filling factor, so that
   their total cross sections cover a fraction of the disk image. This case is reduced to case (3) by
   considering the total cross section of the clumps.}

  \begin{table}[t]\begin{center}
  \begin{tabular}{cc}
   \multicolumn{2}{c}{XL} \\

  \begin{tabular}{|c|ccc|}
  \hline
  3 species &  T25 & DP25 & R25 \\
  \hline
  N      &  1.8~{\rm au} &  2.5~{\rm au} &  2.8~{\rm au} \\
  T25  &               & 2.2~{\rm au} &  1.8~{\rm au} \\
  DP25 &               &                &  2.5~{\rm au} \\
  \hline
  \multicolumn{1}{c}{\phantom{888 species}} &
  \multicolumn{1}{c}{\phantom{888.88}} &
  \multicolumn{1}{c}{\phantom{888.88}} &
  \multicolumn{1}{c}{\phantom{888.88}}
  \end{tabular}

   &

  \begin{tabular}{|c|ccc|}
  \hline
  3 species &  T50 & DP50 & R50 \\
  \hline
  N      &  4.0~{\rm au} &  4.9~{\rm au} &  5.3~{\rm au} \\
  T50  &               & 4.9~{\rm au} &  3.9~{\rm au} \\
  DP50 &               &                &  4.8~{\rm au} \\
  \hline
  \multicolumn{1}{c}{\phantom{888 species}} &
  \multicolumn{1}{c}{\phantom{888.88}} &
  \multicolumn{1}{c}{\phantom{888.88}} &
  \multicolumn{1}{c}{\phantom{888.88}}
  \end{tabular}

   \\
   \multicolumn{2}{c}{XM} \\

  \begin{tabular}{|c|ccc|}
  \hline
  3 species &  T25 & DP25 & R25 \\
  \hline
  N      &  1.6~{\rm au} &  1.7~{\rm au} &  2.7~{\rm au} \\
  T25  &               & 2.5~{\rm au} &  1.7~{\rm au} \\
  DP25 &               &                &  1.9~{\rm au} \\
  \hline
  \multicolumn{1}{c}{\phantom{888 species}} &
  \multicolumn{1}{c}{\phantom{888.88}} &
  \multicolumn{1}{c}{\phantom{888.88}} &
  \multicolumn{1}{c}{\phantom{888.88}}
  \end{tabular}

   &

  \begin{tabular}{|c|ccc|}
  \hline
  3 species &  T50 & DP50 & R50 \\
  \hline
  N      &  3.8~{\rm au} &  4.0~{\rm au} &  5.3~{\rm au} \\
  T50  &               & 5.1~{\rm au} &  3.9~{\rm au} \\
  DP50 &               &                &  4.1~{\rm au} \\
  \hline
  \multicolumn{1}{c}{\phantom{888 species}} &
  \multicolumn{1}{c}{\phantom{888.88}} &
  \multicolumn{1}{c}{\phantom{888.88}} &
  \multicolumn{1}{c}{\phantom{888.88}}
  \end{tabular}

   \\
   \multicolumn{2}{c}{XS} \\

  \begin{tabular}{|c|ccc|}
  \hline
  3 species &  T25 & DP25 & R25 \\
  \hline
  N      &  1.6~{\rm au} &  1.6~{\rm au} &  2.1~{\rm au} \\
  T25  &               & 2.7~{\rm au} &  2.0~{\rm au} \\
  DP25 &               &                &  2.1~{\rm au} \\
  \hline
  \multicolumn{1}{c}{\phantom{888 species}} &
  \multicolumn{1}{c}{\phantom{888.88}} &
  \multicolumn{1}{c}{\phantom{888.88}} &
  \multicolumn{1}{c}{\phantom{888.88}}
  \end{tabular}

   &

  \begin{tabular}{|c|ccc|}
  \hline
  3 species &  T50 & DP50 & R50 \\
  \hline
  N      &  3.8~{\rm au} &  3.8~{\rm au} &  4.3~{\rm au} \\
  T50  &               & 5.7~{\rm au} &  4.7~{\rm au} \\
  DP50 &               &                &  4.8~{\rm au} \\
  \hline
  \multicolumn{1}{c}{\phantom{888 species}} &
  \multicolumn{1}{c}{\phantom{888.88}} &
  \multicolumn{1}{c}{\phantom{888.88}} &
  \multicolumn{1}{c}{\phantom{888.88}}
  \end{tabular}

   \\
  \end{tabular}
  \end{center}\caption{  The upper limits to the sizes of the LMG clumps that exist on
   $25{\mathrm{au}} < r < 50{\mathrm{au}}$ and
   $50{\mathrm{au}} < r < 100{\mathrm{au}}$ orbit, respectively.    }\label{tbl:UpperLimit}
  \end{table}

   We can put the upper limit to the size of such LMG clumps by thresholding the measure-of-sensitivity.
   For example, if the radii of LMG clumps is smaller than the values in Table \ref{tbl:UpperLimit},
   they are not $5.0-\sigma$ detectable.
   The matched filter studies show that the Townsend breakdown model is the easiest model to detect,
   Druyversteyn-Penning model being next, runaway breakdown being most difficult.
   The tendency is explained as the wider the Doppler broadening is, more easier is the detection.
  \section{CONCLUSIONS AND DISCUSSIONS.}
   Discharge phenomena take place in the regions with the critical
   electric field (LMG), and we have established observable features for detecting LMGs by
   the line observations of the accelerated molecular ions.
   Dielectric strength of the disk gas, being one of the crucial elementary processes,
   will open up the understanding of the MRI in protoplanetary disks.
   Understanding of the MRI in weakly-ionized accretion disks will
   contribute to the study of the dynamics of protoplanetary disks
   as well as circumplanetary disks \citep{{bibcode:2014MNRAS.440...89K}}.

   We have presented three dielectric strength models for protoplanetary disks.
   They are Townsend breakdown model, Druyversteyn-Penning breakdown model,
   and runaway breakdown model, respectively.
   We have proposed a method for observational distinguishment of the three models.
   The models are distinguishable with
   the sensitivity of advanced telescopes such as ALMA.
   It is now possible to reject some of the lightning models based on ground observations.
   The upper limits of the LMG clouds size are given from the observations.

   Our lightning models treated here are quite simple. Further studies are targeting to apply this work to more realistic
   disk models as well as more detailed discharge models \citep{{bibcode:2015ApJ...800...47O}}.

   \section*{Acknowledgement}
        We thank Yasuo Fukui, Hitoshi Miura, Munetake Momose and Takayuki Muto for helpful discussions;
        Shinichi Enami for his advices on ice surface charge chemistry;
        Edward Kmett, Simon Peyton Jones and Richard Eisenberg for their comments on our Haskell programs;
        Tooru Eguchi for instruction of Japanese Virtual Observatory;
        Motomichi Tashiro for discussion on empirical formula of ion-molecule collision cross section.
        {We thank Steven Rieder and Keigo Nitadori in helping us correct grammatical errors.}
        We are grateful to
        Institute of Low Temperature Science, Hokkaido University for hosting the workshop
        ``Recent Development in Studies of Protoplanetary Disks with ALMA''
        where the authors learned ALMA image analysis.
        This work is based on the landmark review of terrestrial lightning by \citet{{isbn:9784130627184}}.
        The parallel computation techniques used in this work are based on \citet{{isbn:9781449335946}}.
        We used {\tt units} library { \citep{{doi:10.1145/2633357.2633362}} }to thoroughly check for the
        consistency of physical dimensions and units in this paper.
        This work is supported by Grants-in-Aid for Scientific Research
        (\#23103005,\#24103506,\#25887023,\#26400224) from MEXT.
        This research used computational resources of the K computer provided by the RIKEN Advanced Institute for Computational Science(AICS). We thank RIKEN AICS for the support in conducting this research.
        \change{We thank the referee for the patient and constructive interaction, with which we have made
        a substantial improvement to this paper.}
         \appendix \section{CROSS SECTION MODEL OF ION-NEUTRAL MOLECULAR COLLISION}
     We establish the model of ion-neutral collisional cross sections as functions of collision energy,
in collaboration with Motomichi Tashiro.

In order to compute the equilibrium speed under electric field for ion species
 $\rm HCO^{+}$ , $\rm DCO^{+}$  and $\rm N_2H^{+}$, we need the knowledge of
the energy-dependent cross section of the collisions between the
 $\rm H_2$ and the ion species.
However, no experimental values for such collisions exist.
In general, collisional cross section data for molecular ions and molecules are scarce, due to
the difficulty of setting up such collision experiments.
On the other hand, quantum-mechanical simulation of such collision event
would require upto 1 month per single collision
(Tashiro, private comm.,) and it requires many collision simulations with different
collision parameters to establish a cross section value for one collision energy.
The computational cost prohibits the simulational estimation of the collisional cross section.

Therefore, we construct and use a simple empirical model of
 molecular ions and molecules collisional cross section, following
Tashiro's advice.

There are  collision cross section data
  \citep{{bibcode:1990JPCRD..19..653P,bibcode:1991JPCRD..20..557P}}.
for the following six pairs of  molecular ions and molecules:
${\rm H^{+}}-{\rm H_2}$ ,
${\rm {H_2}^{+}}-{\rm H_2}$ ,
${\rm {H_3}^{+}}-{\rm H_2}$ ,
${\rm {N}^{+}}-{\rm N_2}$ ,
${\rm {N_2}^{+}}-{\rm N_2}$ , and
${\rm {Ar}^{+}}-{\rm Ar}$ .
We use the following model of total collisional cross section
$\sigma_{I^{+}, I}(\varepsilon)$ between molecular ion species
$I^{+}$ and ion species $I$:
\begin{eqnarray}
\sigma_{I^{+}, I}(\varepsilon) = A(\varepsilon) \mu(I^{+}, I)^{p(\varepsilon)},
\label{eq-fit-model}
\end{eqnarray}
where
$\varepsilon$ is the collision energy,
  $\mu(I^{+}, I)$ is the residual mass of species
$I^{+}$ and $I$.
 $A(\varepsilon)$ and
$p(\varepsilon)$ are model parameters
\change{universal across all species}.

\begin{figure}
   \begin{center}
\tabcolsep=0cm
\begin{tabular}{cccc}
\multirow{3}{*}{\rotatebox[origin=c]{90}{collision cross section $(\times 10^{ -16} \mathrm{cm}^2)$ \hspace{3cm}}}
&
\includegraphics[angle=270,width=5cm]{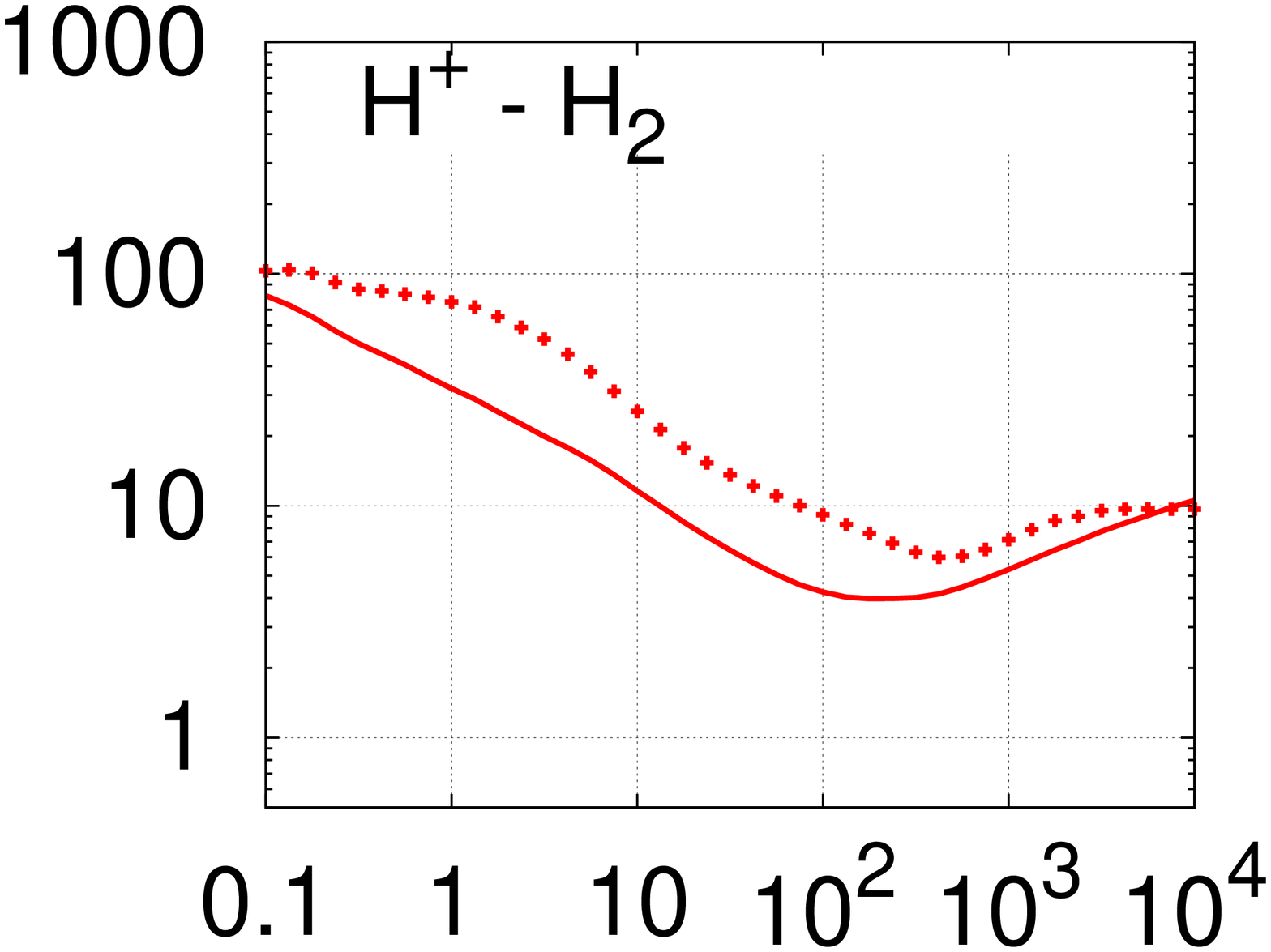}&
\includegraphics[angle=270,width=5cm]{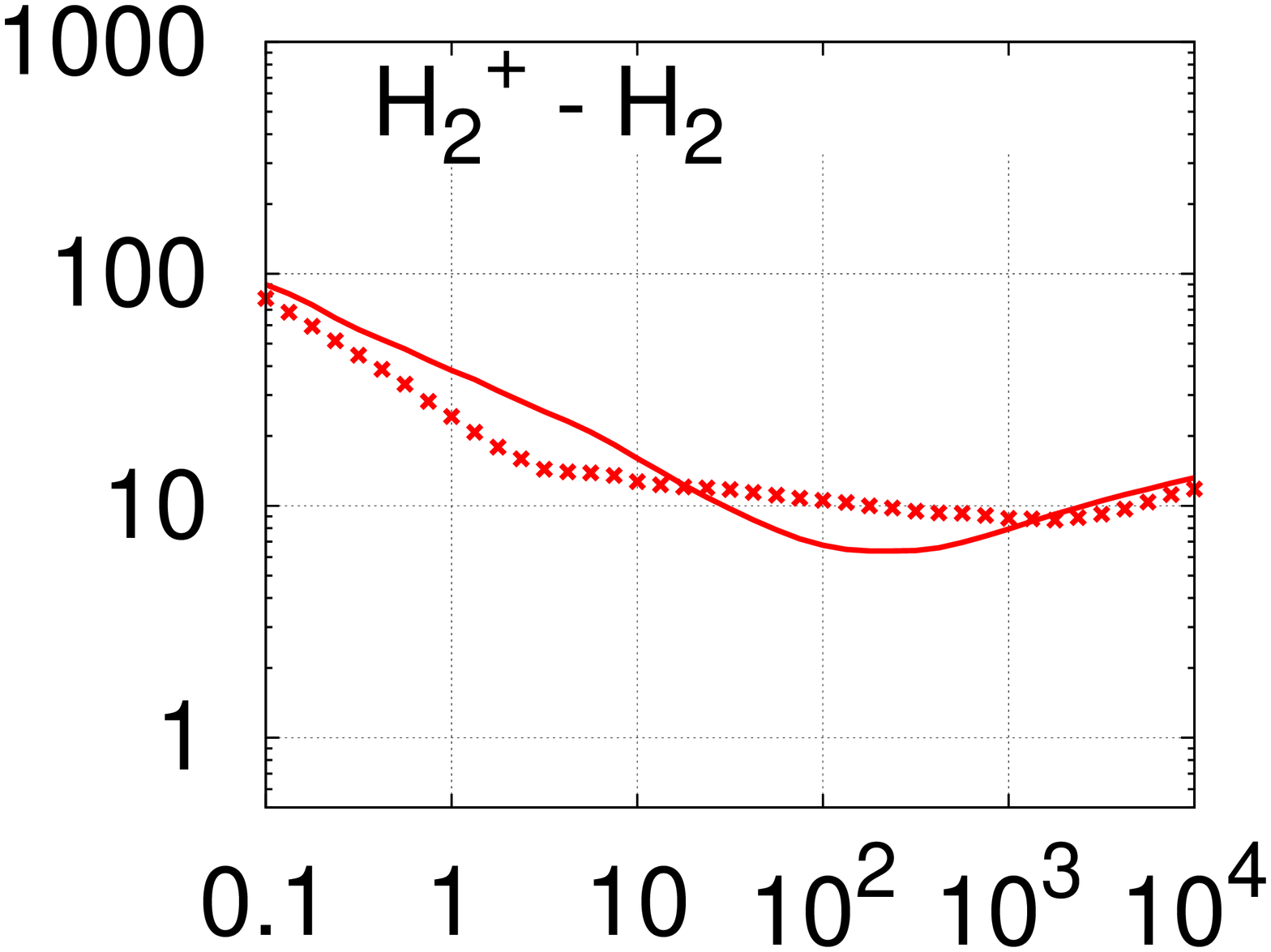}&
\includegraphics[angle=270,width=5cm]{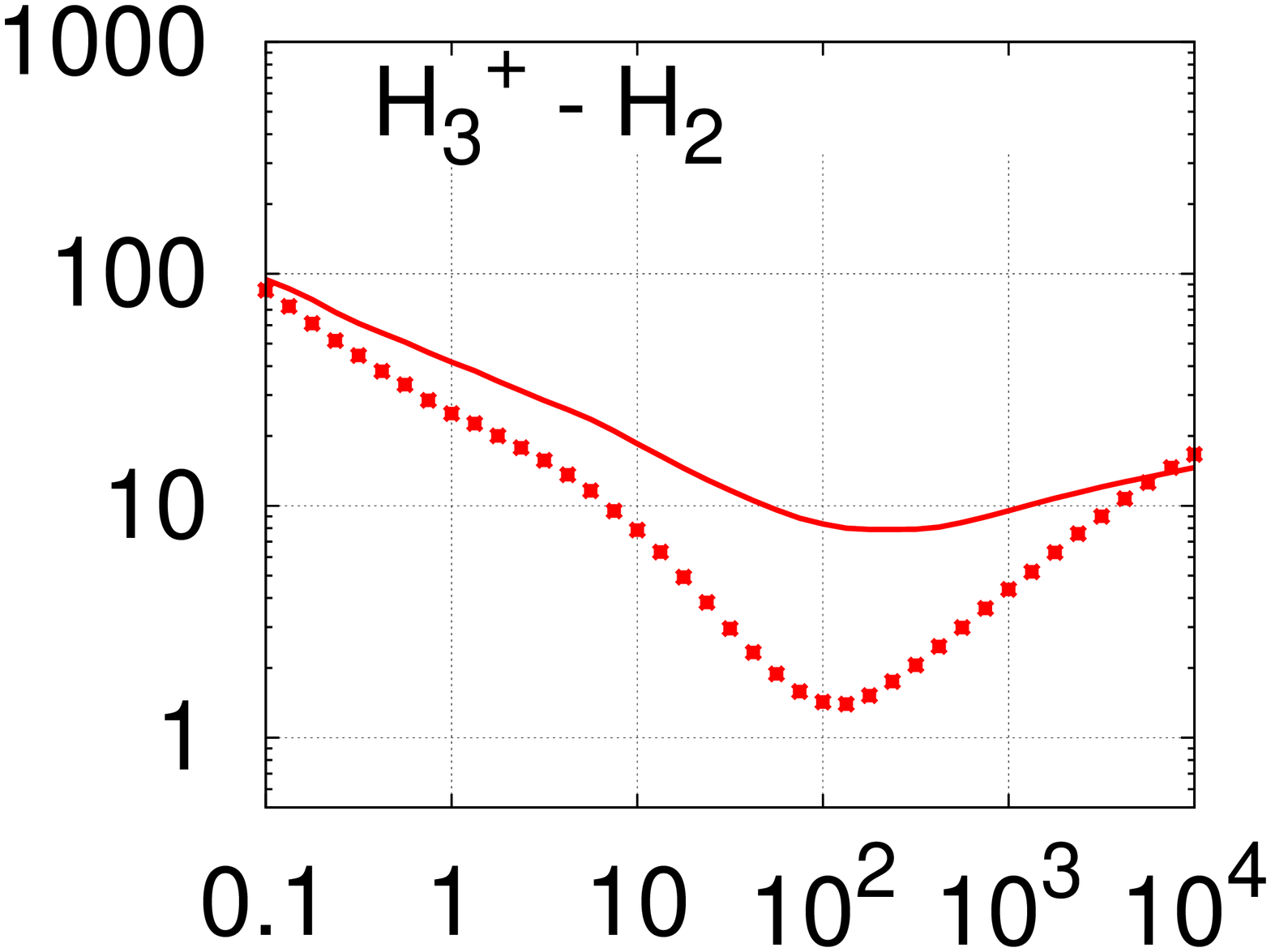}\\
&
\includegraphics[angle=270,width=5cm]{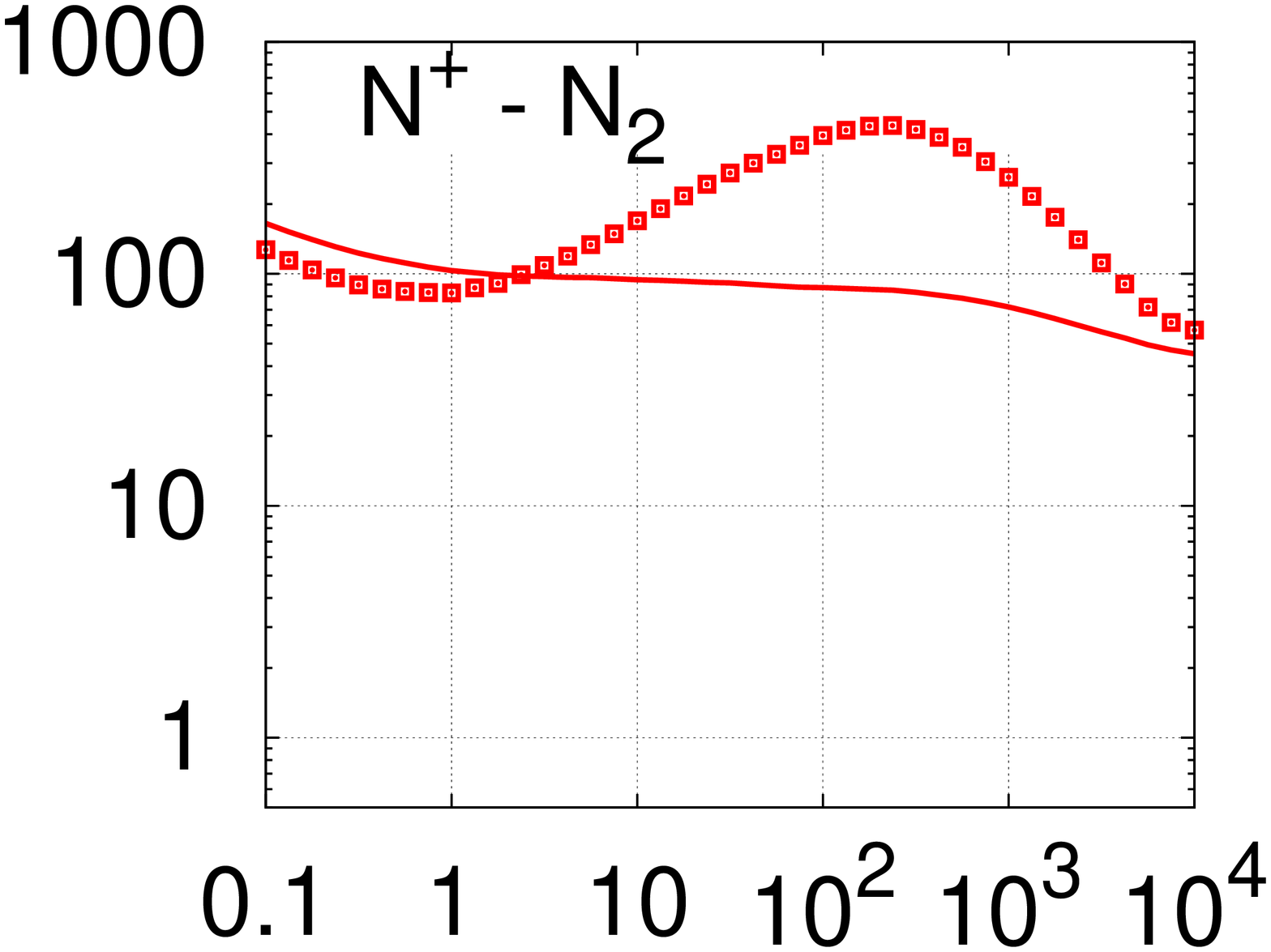}&
\includegraphics[angle=270,width=5cm]{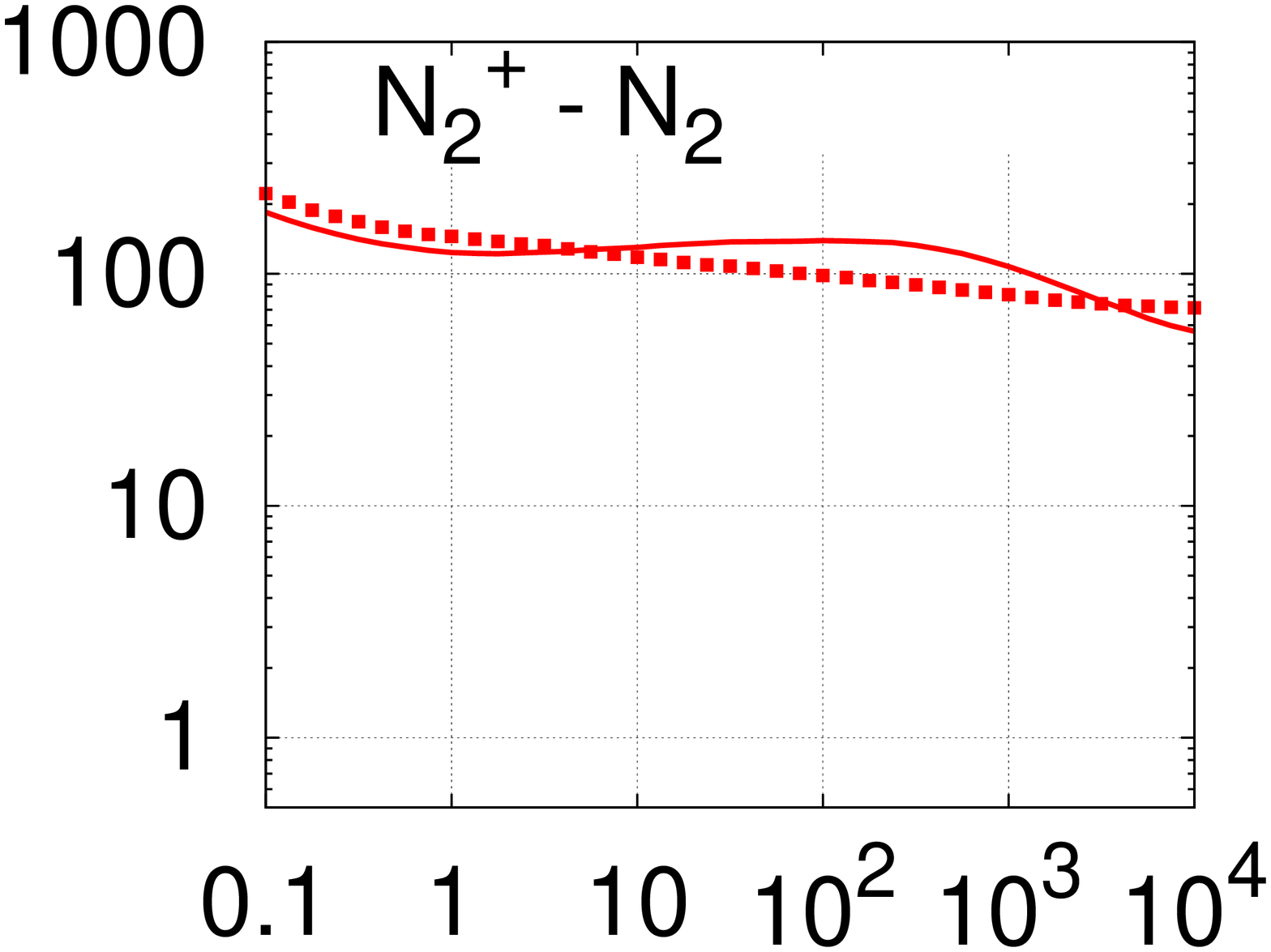}&
\includegraphics[angle=270,width=5cm]{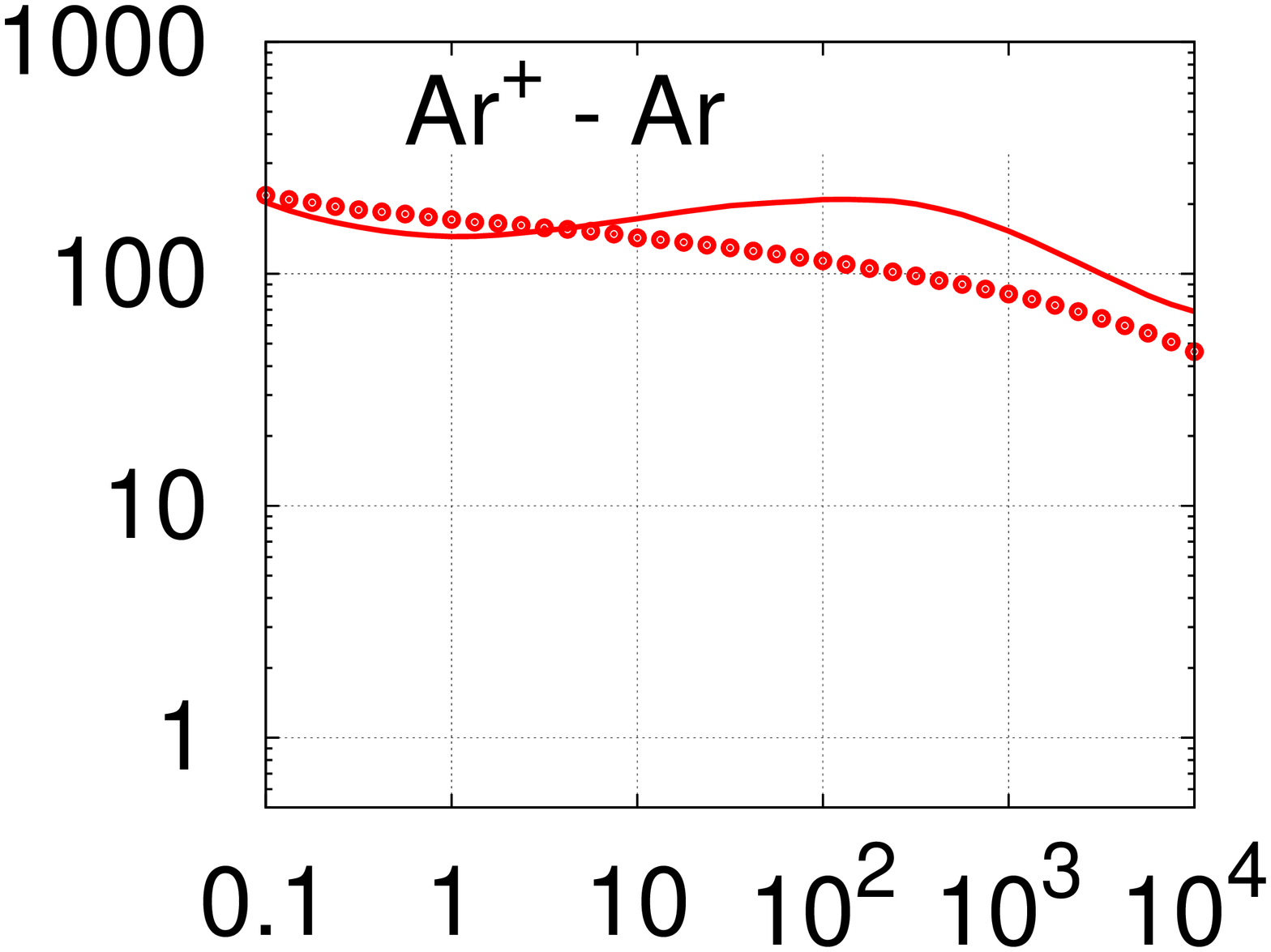}\\
&
\includegraphics[angle=270,width=5cm]{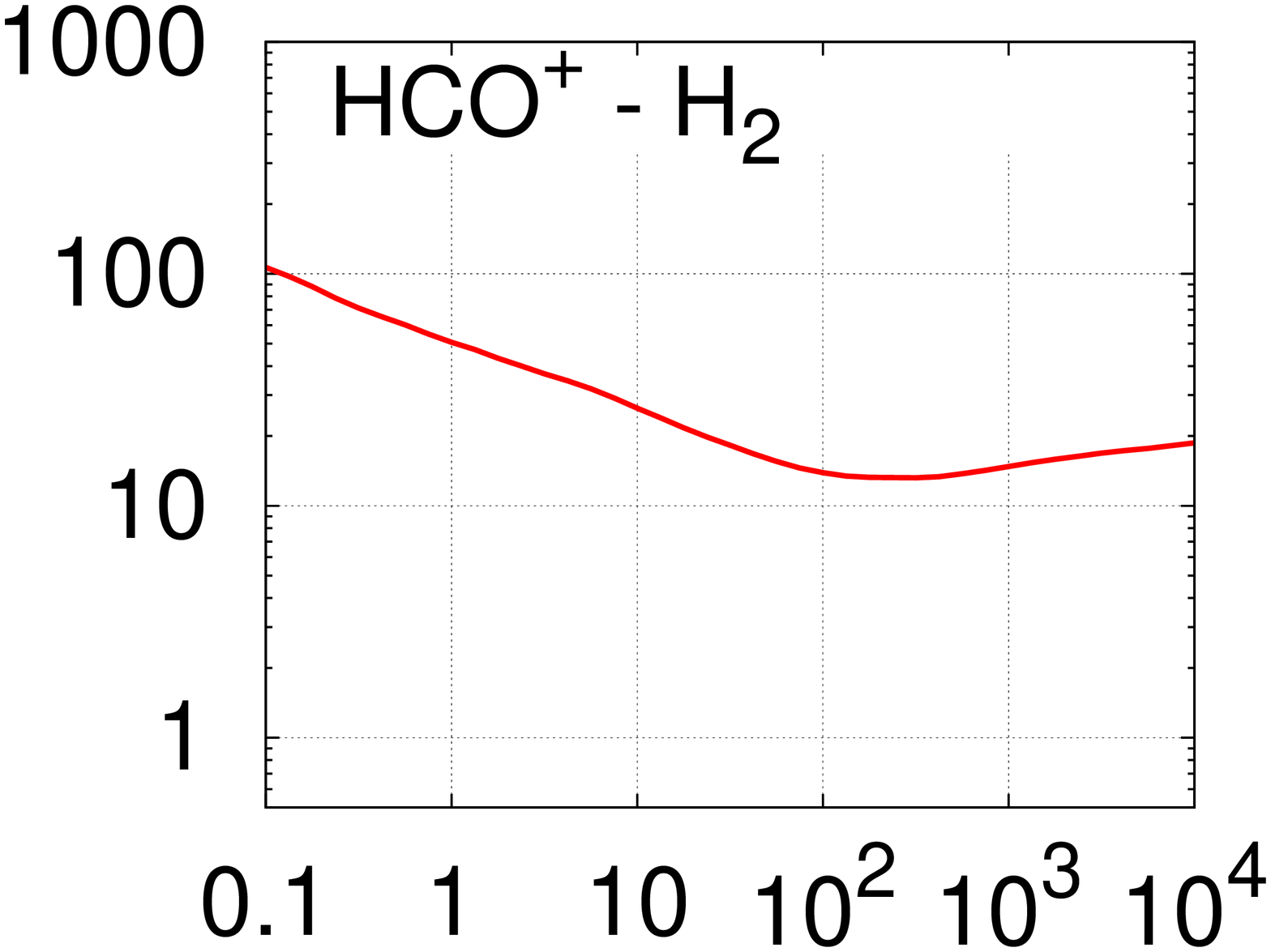}&
\includegraphics[angle=270,width=5cm]{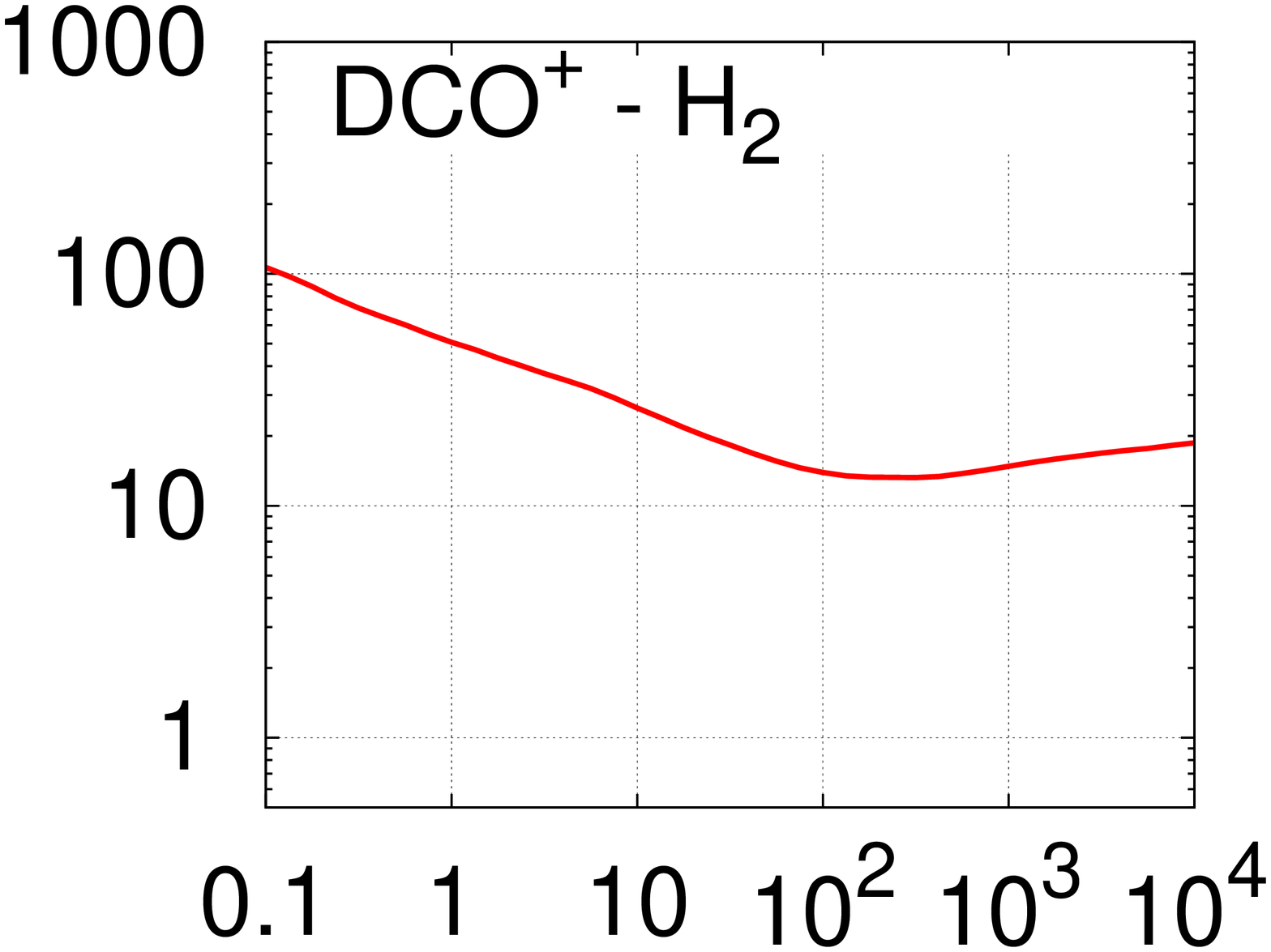}&
\includegraphics[angle=270,width=5cm]{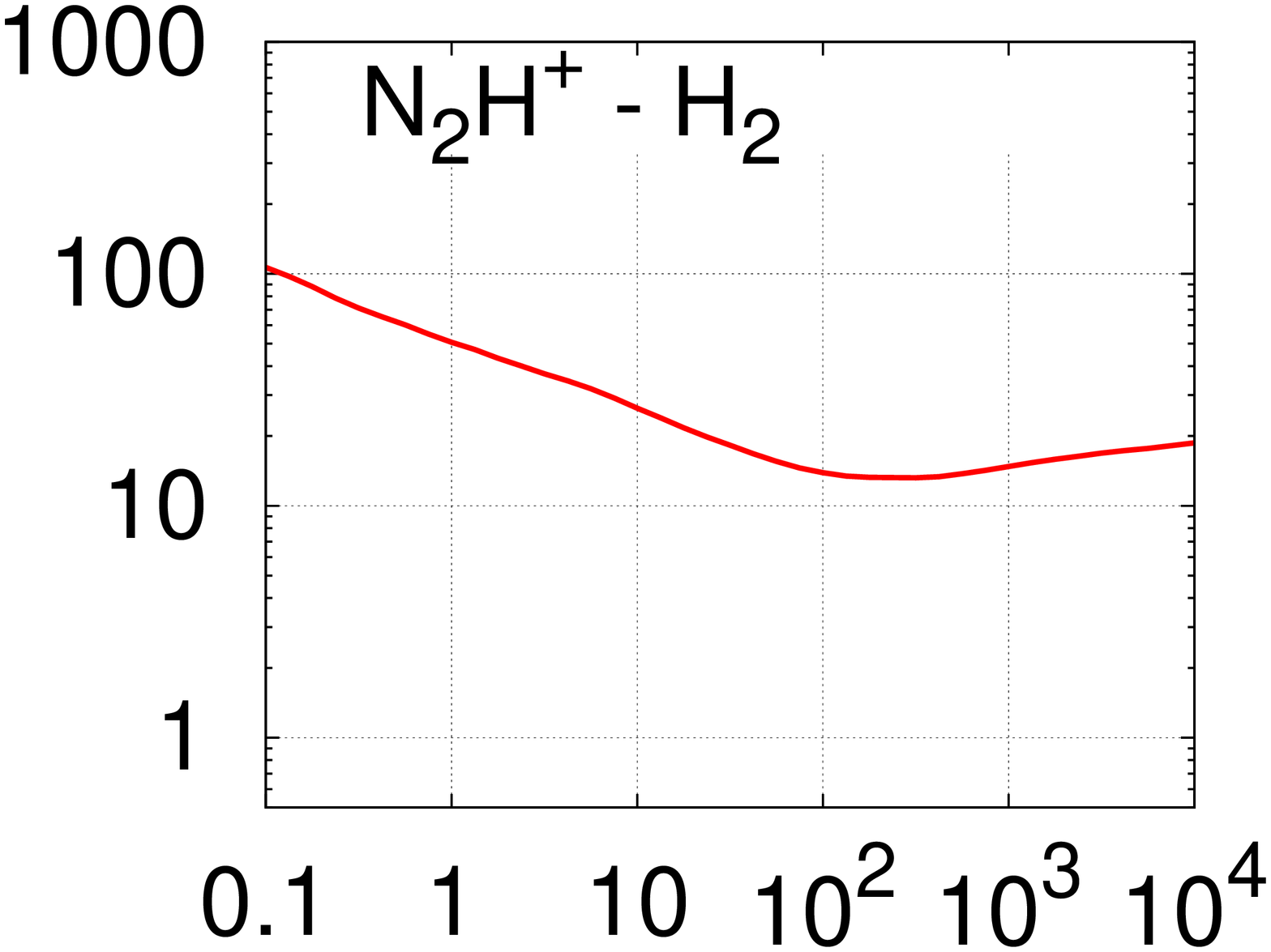}\\
&\multicolumn{3}{c}{collision energy (eV)}
\end{tabular}
   \end{center}
\caption{{Cross section models. Experimental data are points, curves show the model XM.}}
 \label{fig-cross-section-model}
  \end{figure}

We perform the fitting of the model so that the following cost function $C$

\begin{eqnarray}
C = \sum_{\varepsilon,I^{+}, I }
\left(\sigma_{I^{+}, I}(\varepsilon) - \sigma_{I^{+}, I,\rm exp}[\varepsilon]
\right)^2
\end{eqnarray}

is minimized. Here, $\sigma_{I^{+}, I,\rm exp}[\varepsilon]$
are experimentally known cross section values for some fixed values of $\varepsilon$
found in
  \citet{{bibcode:1990JPCRD..19..653P,bibcode:1991JPCRD..20..557P}}.
The experimental data and the best-fit models are presented in
Figure \ref{fig-cross-section-model}. Our cross section model predicts almost identical cross sections for
the three collision pairs $\rm HCO^{+}-H_2$ , $\rm DCO^{+}-H_2$,  and $\rm N_2H^{+}-H_2$.
This is because our model depends only on
the residual mass of species $\mu(I^{+}, I)$, and all the three pairs
 $\rm HCO^{+}-H_2$ , $\rm DCO^{+}-H_2$,  and $\rm N_2H^{+}-H_2$ have similar residual masses.

   {
Available experimental data on ion-neutral collisional cross section are scarce,
and do not justify
 models more complex than equation (\ref{eq-fit-model}).
Instead of making better-fitting models, we study how the uncertainty in the model affects
our results. In addition to the  model equation
(\ref{eq-fit-model}) with the fitted parameters
 (Figure \ref{fig-cross-section-model}), we study two alternative models, where
\begin{eqnarray}
\sigma_{I^{+}, I}^{\rm L}(\varepsilon) &=& 10 \sigma_{I^{+}, I}(\varepsilon) , \label{eq-hi-model} \\
\sigma_{I^{+}, I}^{\rm S}(\varepsilon) &=& \frac{1}{10} \sigma_{I^{+}, I}(\varepsilon) , \label{eq-lo-model}
\end{eqnarray}
for $I^{+}-I = \rm HCO^{+}-H_2$ , $\rm DCO^{+}-H_2$,  and $\rm N_2H^{+}-H_2$ .
The cross section models of
equations
(\ref{eq-fit-model}),
(\ref{eq-hi-model}), and
(\ref{eq-lo-model}),
are labeled
XM, XL, and XS,
respectively.
}

%
% These complex models reproduce more detailed behavior of the cross sections such as
% dependence on ion-neutral molecular mass ratio.
% However, there are numerous candidates for such models, and the
% predicted cross section values vary upto four degrees of magnitude among the models.
% Available experimental data are not abundant enough to justify a choice
% among those complex models without risking overfitting.
% Therefore, we decided to use the simple model of equation (\ref{eq-fit-model})
% that captures overall tendency.

\end{document}